\documentclass[aps,nofootinbib,preprintnumbers,tightenlines,twocolumn,superscriptaddress,longbibliography]{revtex4-1}
\usepackage{braket}
\usepackage{array}
\usepackage{dsfont}
\usepackage{amsmath}
\usepackage{pifont}
\usepackage[normalem]{ulem}
\usepackage{amssymb}
\usepackage{xcolor}
\usepackage{amsfonts}
\usepackage{mathtools}
\usepackage{graphicx}
\usepackage{dcolumn}
\usepackage{bm}
\usepackage{multirow}
\usepackage{hhline}

\usepackage{leftidx}
\usepackage{color}
\usepackage{mathtools}
\usepackage{MnSymbol}
\usepackage[mathscr]{eucal}
\usepackage[capitalize]{cleveref}

\usepackage[utf8]{inputenc}
\usepackage{natbib}
\usepackage{comment}
\usepackage{xcolor}
\newcommand{\abs}[1]{\left| #1 \right|}

\usepackage[normalem]{ulem}

\begin{document}

\title{Interferometry of Efimov states in thermal gases by modulated magnetic fields}

\author{G. Bougas}
\email{gbougas@physnet.uni-hamburg.de}
\affiliation{Center for Optical Quantum Technologies, Department of Physics, University of Hamburg, Luruper Chaussee 149, 22761 Hamburg Germany }

\author{S. I. Mistakidis}
\affiliation{ ITAMP, Center for Astrophysics $\vert$ Harvard \& Smithsonian, Cambridge, MA 02138 USA}
\affiliation{Department of Physics, Harvard University, Cambridge, Massachusetts 02138, USA}

\author{P. Schmelcher}
\affiliation{Center for Optical Quantum Technologies, Department of Physics, University of Hamburg, Luruper Chaussee 149, 22761 Hamburg Germany }
\affiliation{The Hamburg Centre for Ultrafast Imaging, University of Hamburg, Luruper Chaussee 149, 22761 Hamburg, Germany}

\author{C. H. Greene}
\affiliation{Department of Physics and Astronomy, Purdue University, West Lafayette, Indiana 47907, USA}
\affiliation{Purdue Quantum Science and Engineering Institute, Purdue University, West Lafayette, Indiana 47907, USA}

\author{P. Giannakeas}
\affiliation{Max-Planck-Institut f\"ur Physik komplexer Systeme, N\"othnitzer Str.\ 38, D-01187 Dresden, Germany }

\begin{abstract}
  We demonstrate that an interferometer based on modulated magnetic field pulses enables precise characterization of the energies and lifetimes of Efimov trimers irrespective of the magnitude and sign of the interactions in $^{85}$Rb thermal gases.
  Despite thermal effects, interference fringes develop when the dark time between the pulses is varied. This enables the selective excitation of coherent superpositions of trimer, dimer and free atom states.
  The interference patterns possess two distinct damping timescales at short and long dark times that are either equal to or twice as long as the lifetime of Efimov trimers, respectively.
  Specifically, this behavior at long dark times provides an interpretation of the unusually large damping timescales reported in a recent experiment with $^7$Li thermal gases [Yudkin \textit{et al}., Phys. Rev. Lett. \textbf{122}, 200402 (2019)].
  Apart from that, our results constitute a stepping stone towards a high precision few-body state interferometry for dense quantum gases.
\end{abstract}

\maketitle

\section{Introduction}  \label{Sec:Introduction}

Efimovian trimers constitute an infinite set of particle triplets occurring in the absence of two-body binding
\cite{Efimov_energy_1970,Efimov_energy_1973,Efimov_weakly_1971,greene_universal_2017,nielsen_three-body_2001,naidon_efimov_2017,dincao_few-body_2018}.
Owing to their universal character, they have been explored in both nuclear and atomic physics \cite{kraemer_evidence_2006,Kunitski_observation_2015,Endo_universal_2016,greene_universal_2017,Kievsky_efimov_2021} and in the context of many-body physics as  
the binding mechanism for magnons \cite{nishida2013efimov} and polaritons \cite{gullans2017efimov}.
Furthermore, the role of Efimov states is pivotal for some ultracold gases in equilibrium, e.g. polarons \cite{tran_fermions_2021,Christianen_bose_2022,Naidon_two_2018,Sun_efimov_2019} and in some out-of-equilibrium \cite{Musolino_bose_2022,Colussi_dynamical_2018,makotyn_universal_2014,eigen_universal_2018,klauss_observation_2017,fletcher_two-_2017}, despite their short lifetime due to collisional decay, i.e. three-body recombination processes.
Recent investigations in dense gas mixtures demonstrate that such processes can be suppressed due to medium effects \cite{blochprl2022supress}.
Specifically, this puts forward the idea that the intrinsic properties of Efimov states , i.e. the binding energies and lifetimes, are potentially modified.
Hence, dynamically probing simultaneously both intrinsic properties of Efimov trimers could provide alternative ways to study the impact of an environment.  

To address such effects, a promising dynamical protocol is to expose a many-body system in a double sequence of magnetic field modulations (pulses).
The latter has been used successfully to precisely measure the binding energies and lifetimes of dimers \cite{donley_atommolecule_2002} near a Feshbach resonance \cite{chin_feshbach_2010}.
Beyond two-body physics, employing this Ramsey-type protocol for a thermal gas of $^7$Li atoms, Yudkin {\it et al.} precisely 
probed Efimov molecules even near the atom-dimer threshold \cite{yudkin_coherent_2019, yudkin_efimov_2020}; an experimentally challenging region.
Specifically, the surviving atom number exhibited damped Ramsey fringes that  
were robust against thermal effects. 
However, the corresponding damping timescale was found to exceed the typical lifetime of Efimov trimers even for $^{85}$Rb$_3$ \cite{klauss_observation_2017}. 
In this regard, it has remained elusive how the lifetime of Efimov trimers emerges in the interference fringes induced by magnetic field pulses.
To address the intricate dynamics of a three-body system requires a time-dependent theoretical framework establishing also a systematic pathway to explore the role of few-body physics in out-of-equilibrium many-body systems \cite{klauss_observation_2017,makotyn_universal_2014}. 

\begin{figure*}[t!]
\centering
\includegraphics[width=0.95 \textwidth]{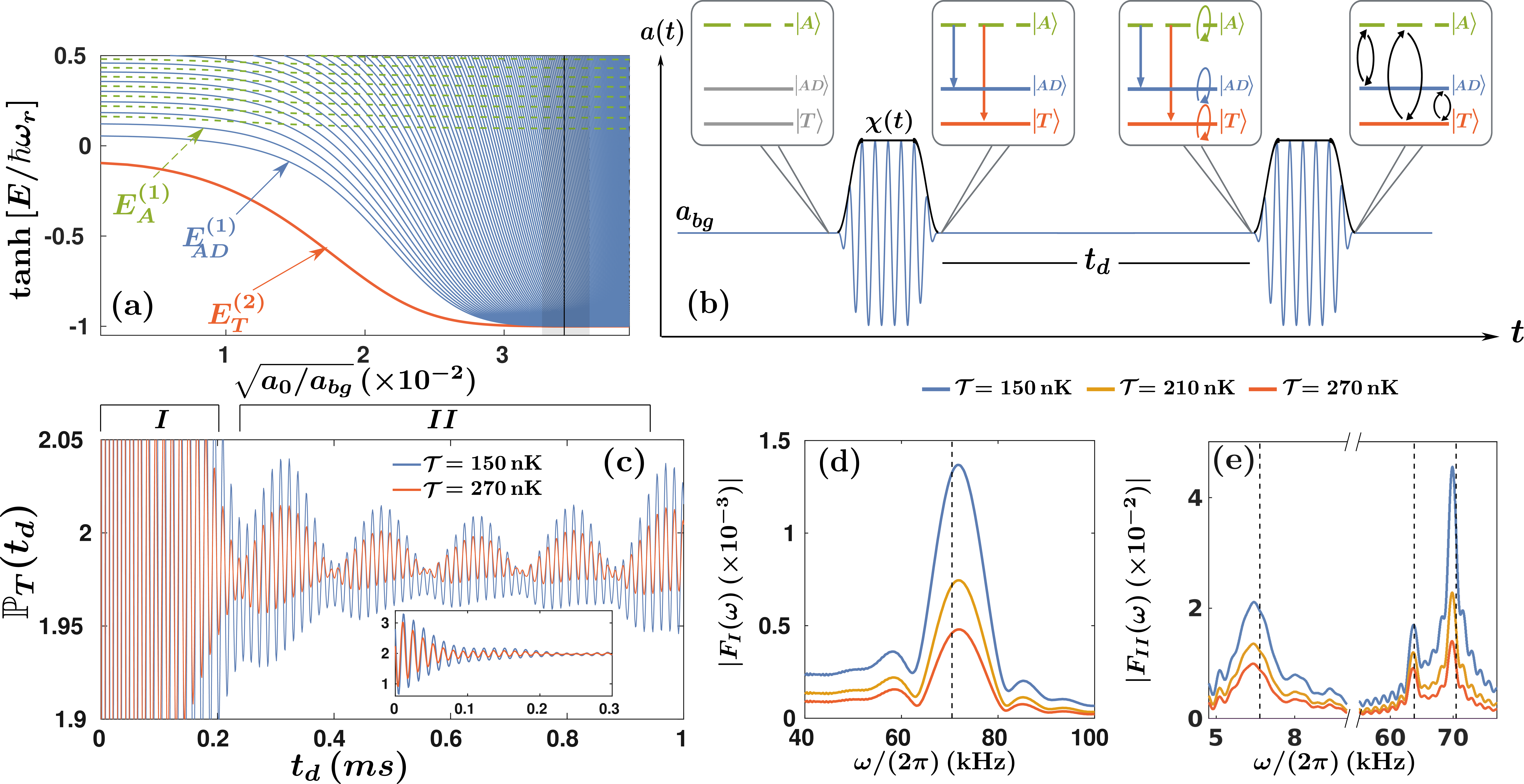}    
\caption{(a) Energy spectrum of three harmonically trapped $^{85}$Rb particles with $\omega_r/(2\pi)= 350 \, \rm{Hz}$.
Efimov trimer (T), atom-dimer (AD) and trap (A) states are depicted.
Initially the scattering length is set at $a_{bg}=819 \, a_0$ (dashed vertical line), then modulated with amplitude $a_m$ (gray region). Note $a_0$ is the Bohr radius.
(b) A schematic illustration of the Ramsey-type interferometer: A first pulse with envelope $\chi(t)$ associates atom-dimers and Efimov states out of trap states (first and second sub-graphs in (b)), the system then evolves freely during the dark time $t_d$ (third sub-graph in (b)), while a second pulse further admixes the states together with their dynamical phases that were accumulated during $t_d$ (fourth sub-graph in (b)).
(c) The ratio of thermally averaged (RTA) probabilities, $\mathbb{P}_{T}(t_d)$ at $a_{bg}=819 \, a_0$ and distinct temperatures (see legend). 
Inset: A zoom out plot of RTA at early $t_d$. 
(d) [(e)] Frequency spectra referring to region I [II] of the RTA quantifying its single [multifrequency] behavior at different values of temperature $\mathcal{T}$. 
The vertical dotted lines correspond to the three-level model (TLM) predictions for $E^{(2)}_{T}$, $E^{(1)}_{AD}$ and a trap state (see text). 
}
\label{Fig:Spectrum}
\end{figure*}

Such an approach is developed here to investigate the three-body dynamics of a thermal gas.
We consider $^{85}$Rb atoms since the lifetimes of the ensuing trimers and dimers are known experimentally \cite{klauss_observation_2017} in contrast to $^7$Li \cite{yudkin_coherent_2019}. 
Our study establishes that, by implementing double magnetic field pulses, the intrinsic properties of Efimov trimers are readily probed regardless of the sign or magnitude of the scattering length; at which these states occur. 
Rich interferometric spectra exhibit both low- and high-frequencies independent of the gas temperature. 
The low-frequency components originate from the coherent superposition of the trimer with the dimer state, consistent with the observations in Ref. \cite{yudkin_coherent_2019}.
The additional high-frequencies arise from the coherent population of the trimer or dimer states with the ones lying at the ``at break-up" threshold.
The characteristic damping time of the field generated interference fringes is shown to be twice the lifetime of the Efimov trimers, providing an explanation for the unusually long decay times observed in Ref. \cite{yudkin_coherent_2019}.

This work begins by introducing the time-dependent framework for the three-body system in Sec. \ref{Sec:Interferometry}, providing also details on the employed techniques. 
Subsequently, in Sec. \ref{Sec:Dynamical}, the association mechanisms of the dynamical scheme are examined. The role of the lifetime of the Efimov states is studied in Sec. \ref{Sec:Lifetime} for both repulsive and attractive background interactions. Sec. \ref{Sec:Conclusions} summarizes our major findings and future perspectives are discussed.
Appendix~\ref{Ap:Three_body} outlines the steps to numerically solve the three-body time-dependent Schr\"odinger equation in hyperspherical coordinates, while the explicit form of the interaction potential matrix elements using field-free eigenstates is given in Appendix~\ref{Ap:Matrix_elements}. 
Further insights into the three-level model via first-order time-dependent perturbation theory are provided in Appendix~\ref{Ap:Three_level}.

\section{Time-dependent three-body system and interferometry protocol} \label{Sec:Interferometry}

Our paradigm system consists of three $^{85}$Rb atoms of mass $m$ confined in a spherically symmetric harmonic trap with radial frequency $\omega_r$. 
Following the prescription of Refs. \cite{Goral_adiabatic_2004,sykes_quenching_2014,corson_bound-state_2015,borca_two-atom_2003,dincao_efimov_2018,von_stecher_spectrum_2007}, we set $\omega_r=2\pi \times 350 \rm{Hz}$ yielding a single atom trap length $a_{r}=\sqrt{\hbar/(m \omega_r)}$, that compares to the interparticle spacing ($\sim \braket{n}^{-1/3}$) used in Ref. \cite{klauss_observation_2017} for a local peak density $n_0=5 \cdot 10^{12} \, \rm{cm}^{-3}$.
The dynamics and the universal characteristics of the three-body system are addressed by employing contact interactions with a time-dependent $s$-wave scattering length, i.e. $a(t)$.
The three-body Hamiltonian reads:
\begin{gather}
\mathcal{H}(t)=\sum_{i=1}^{3} \left( \frac{-\hbar^2  \nabla_i^2}{2m}  +\frac{m\omega_r^2}{2} \boldsymbol{r}_i^2 \right)  +\sum_{i<j} \frac{4\pi \hbar^2 a(t)}{m} \delta(\boldsymbol{r}_{ij})\hat{O}_{ij},
\label{Eq:hamilt_lab}
\end{gather}
where $\boldsymbol{r_i}$ denotes the position of the $i$-th atom, and $\hat{O}_{ij}=\partial_{r_{ij}}(r_{ij}\cdot)$ is the Fermi-Huang regularization operator with ${r}_{ij}=|{\boldsymbol{r}}_i-{\boldsymbol{r}}_j|$. 
\cref{Fig:Spectrum}(b) depicts the dynamical profile of $a(t)$ determined by the double pulse magnetic field sequence used in Ref. \cite{yudkin_coherent_2019}, namely
\begin{gather}
a(t)=  a_{bg}+a_m \cos{(\Omega t)} \Big[   \chi(t)+\chi(t-t_d-2t_0-\tau)  \Big], \label{Eq:Pulse} \\
\rm{with}~~\chi(t)= \begin{cases}   \sin^2\left( \frac{\pi t}{2t_0} \right), & 0\leq t < t_0\\
1, & t_0 \leq t < t_0+\tau \\
\sin^2\left( \frac{\pi (t-\tau)}{2t_0} \right), & t_0+\tau \leq t \leq 2t_0+\tau \\
0, & \rm{otherwise}
\end{cases}. \label{Eq:Pulse_envelope}
\end{gather}
Here, $a_{bg}$ indicates the background scattering length of the time-independent system, and $a_m$ is the pulse's amplitude yielding 
$\sim 20 \%$ change to $a_{bg}$. 
$\Omega$ is the driving frequency and $\chi(t)$ denotes the envelope of the pulse where $t_0$ and $\tau$ are the ramp on/off times and length of the pulse envelope, respectively. 
The time between the two pulses is represented by $t_d$, i.e. {\it dark time}, where the system freely evolves.

Owing to \cref{Eq:Pulse}, it suffices to simulate the corresponding time-dependent Schr\"odinger equation in the center-of-mass of the three-body system. Namely, only the relative Hamiltonian depends explicitly on time, i.e. $\mathcal{H}(t) = \mathcal{H}_{cm} + \mathcal{H}_{rel}(t)$. Regarding the center-of-mass Hamiltonian $\mathcal{H}_{cm}$ we assume from here on that the three-atom setting always resides in its ground state, $\ket{0}_{cm}$. 
Subsequently, the $\mathcal{H}_{rel}(t)$ is further decomposed into two terms: (i) a field-free Hamiltonian that describes three atoms in a spherical trap interacting with $a_{bg}$ scattering length and (ii) an explicit time-dependent interaction term [for more details see \cref{Ap:Three_body_decomposition}].
The spectrum of the relative field-free Hamiltonian is obtained via the adiabatic hyperspherical approach \cite{greene_universal_2017,naidon_efimov_2017,dincao_few-body_2018,nielsen_three-body_2001,rittenhouse_greens_2010}. In this method all the relative degrees of freedom are expressed by a hyperradius $R$, that describes the overall system size, and a set of five hyperangles $\boldsymbol{\varpi}$, that address the relative particle positions. Subsequently, the field-free eigenstates $\ket{n}$ are expanded in a set of hyperangular basis functions, $\Phi_{\nu}(R;\boldsymbol{\varpi})$, treating the hyperradius as an adiabatic parameter \cite{greene_universal_2017,nielsen_three-body_2001},

\begin{equation}
       \braket{R,\boldsymbol{\varpi} | n} = R^{-5/2} \sum_{\nu} F^{(n)}_{\nu}(R) \Phi_{\nu}(R;\boldsymbol{\varpi}),
       \label{Eq:Adiabatic_expansion}
\end{equation} 
where the expansion coefficients $F^{(n)}_{\nu}(R)$ are the so-called hyperradial channel functions.

Within the  adiabatic hyperspherical approach, the determination of the eigenstates $\ket{n}$ along with their corresponding eigenenergies is performed in two steps. The hyperangular wavefunctions are obtained first, treating $R$ as an adiabatic parameter. Subsequently, the hyperradial channel functions and $E^{(n)}$ are calculated from the resulting equations that include all the relevant nonadiabatic coupling terms. A more elaborate discussion on the adiabatic hyperspherical approach is provided in \cref{Ap:Three_body_background}.

The stationary eigenenergies $E^{(n)}$ versus the scattering length $a_{bg}$ are shown in \cref{Fig:Spectrum}(a). 
Their corresponding eigenstates, $\ket{n}$, fall into three classes: Efimov trimers ($T$), atom-dimers ($AD$) and trap ($A$) states [red, blue and green lines in  \cref{Fig:Spectrum} (a)].
Furthermore, the adiabatic hyperspherical approach allows to express the time-dependent wave function of \cref{Eq:hamilt_lab} in terms of the field-free eigenstates, i.e. $\ket{\Psi^{(\alpha)}_{3b}(t)}=\sum_n c^{(\alpha)}_n(t) \ket{n} \ket{0}_{cm}$ with $c^{(\alpha)}_n(t)$ being the probability amplitude of the $n$-th stationary state. 
The initial boundary condition is $c^{(\alpha)}_n(0)=\delta_{n \alpha}$ where the index $\alpha$ enumerates solely trap states, i.e. $\alpha \in A$.  
Plugging this expansion into the time-dependent Schr\"odinger equation (TDSE) under the Hamiltonian of Eq. \eqref{Eq:hamilt_lab} leads to a matrix differential equation for the time-dependent expansion coefficients,

\begin{equation}
      i \hbar \frac{d \boldsymbol{c}^{(\alpha)}(t)}{dt} = \boldsymbol{\mathcal{H}}_{rel}(t) \cdot \boldsymbol{c}^{(\alpha)}(t).
      \label{Eq:Expansion_coefficients}
\end{equation}
Here, $\boldsymbol{\mathcal{H}}_{rel}(t)$ represents the relative Hamiltonian matrix expressed in the field-free basis. Given the decomposition of $\mathcal{H}_{rel}(t)$ into a field-free Hamiltonian and an explicit time-dependent interaction term, it is convenient to employ the second-order split-operator method \cite{Burstein_third_1970,Tarana_femtosecond_2012} for solving \cref{Eq:Expansion_coefficients} [for additional information refer also to \cref{Ap:Three_body_propagation}].

According to \cref{Fig:Spectrum}(b), initially the three particles interact with $a(t=0)=a_{bg}=819~a_0$ [see dashed vertical line in \cref{Fig:Spectrum}(a)] residing in a specific trap state.
Similar to Ref. \cite{yudkin_coherent_2019}, at $a_{bg}$ the system supports two Efimov trimer states, with the second (excited) one at energy $E_{T}^{(2)}$  lying close to the first atom-dimer energy in the trap, $E_{AD}^{(1)}$, which represents the atom-dimer threshold. 
At $t\neq0$ the first pulse turns on with an envelope $\chi(t)$ of amplitude $a_m$ [gray region in \cref{Fig:Spectrum} (a)], where $a(t)$ modulates with angular frequency $\Omega$ \cite{giannakeas_nonadiabatic_2019,yudkin_coherent_2019}.
The latter is equal to the energy difference between the first trap and atom-dimer states, i.e. $\Omega/2\pi = (E^{(1)}_A-E_{AD}^{(1)})/h=63.8 ~ \rm{kHz}$, as in the experiment of Ref. \cite{yudkin_coherent_2019}.
Furthermore, the pulse's full-width-at-half-maximum is $27 \,{\rm \mu s}$ providing an energy bandwidth of $6.5 ~ \rm{kHz}$ matching the energy difference between the second trimer and first atom-dimer states, $|E_T^{(2)}-E_{AD}^{(1)}|/h$.
This implies that the first excited trimer $E_T^{(2)}$ and atom-dimer $E_{AD}^{(1)}$ states are coherently populated since the pulse cannot energetically resolve them. 
After the first pulse, the system occupies several $\ket{n}$ eigenstates which freely evolve during the dark time $t_d$, each accumulating a dynamic phase [see \cref{Fig:Spectrum}(b)].
At $t=t_d$, a second pulse, identical to the first one, is applied, admixing different stationary eigenstates and their corresponding dynamical phases.
By the end of the second pulse, we extract the probability to occupy the Efimov trimer state as a function $t_d$. 

In a typical experiment, the three-body dynamics takes place in a thermal gas at temperature $\mathcal{T}$  
\cite{yudkin_coherent_2019,yudkin_efimov_2020}. 
Hence, after the double pulse sequence the probability density to occupy the Efimov trimer needs to be thermally averaged over a Maxwell-Boltzmann ensemble of initial trap states.
For our purposes, we introduce a ratio of thermally averaged (RTA) probabilities, $\mathbb{P}_T(t_d)$, to populate Efimov trimer states after two pulses (numerator) versus one pulse (denominator),
\begin{subequations}
\begin{gather}
       \mathbb{P}_{T}(t_d)=\frac{\sum_{\alpha \in A} \sum_{j \in \, T} e^{-\frac{E_A^{(\alpha)}}{k_B \mathcal{T}}} \abs{c^{(\alpha)}_j(2\tilde{\tau}+t_d)}^2}{\sum_{\alpha \in A} \sum_{j \in \, T}  e^{-\frac{E_A^{(\alpha)}}{k_B \mathcal{T}}} \abs{c^{(\alpha)}_j(\tilde{\tau})}^2 },  \label{Eq:Efimov_fraction} \\
       c^{(\alpha)}_j(2\tilde{\tau}+t_d)=\sum_{n} U_{jn}(2\tilde{\tau}+t_d,\tilde{\tau}+t_d) e^{-i E^{(n)}t_d/\hbar} U_{n \alpha}(\tilde{\tau},0),
       \label{Eq:Evolution_op}
\end{gather}
\end{subequations}
where $k_B$ is the Boltzmann constant, $\tilde{\tau}=2t_0+\tau$ is the pulse duration, and $U_{ij}(\cdot,\cdot)$ represents the three-body evolution operator during a single pulse, expressed in the field-free basis.

\section{Dynamical superposition of Efimov trimers}  \label{Sec:Dynamical}

\begin{figure}[t!]
\centering
\includegraphics[width=0.48\textwidth]{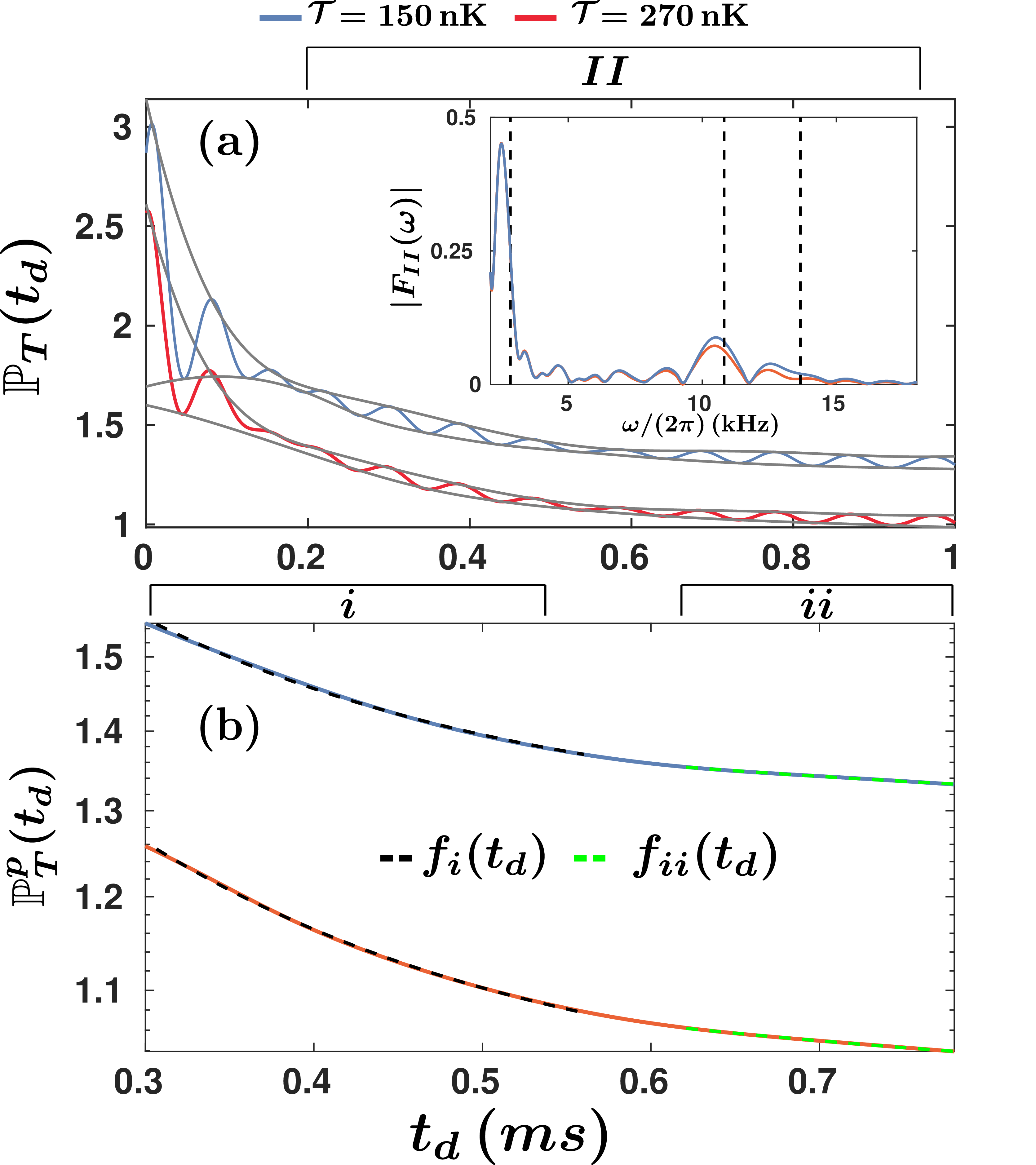}
\caption{(a) $\mathbb{P}_{T}(t_d)$ for different temperatures (see legend), taking into account the decay width, $\Gamma^{(2)}/h=748 \, \rm{Hz}$ of the first excited Efimov state at $a_{bg}=2030 \, a_0$. 
The gray solid lines outline the upper and lower peak envelopes. 
The inset presents the frequency spectrum pertaining to region II, $\abs{F_{II}(\omega)}$. 
(b) The mean peak-to-peak envelope, $\mathbb{P}^p_{T}(t_d)$ is fitted with the exponentials $f_{i/ii}(t_d)=g_{i/ii}e^{-\Gamma_{i/ii}(t_d-t^0_{i/ii})/\hbar}+w_{i/ii}$ at dark time intervals $i$ and $ii$ (black and green dashed lines) with $g_{i/ii}$, $w_{i/ii}$ representing fitting constants. 
The characteristic decay time of the oscillations at long $t_d$ is twice as long as the intrinsic Efimov lifetime $\hbar/\Gamma^{(2)}$.}
\label{Fig:Decay_abg_2030a0}
\end{figure}

\cref{Fig:Spectrum}(c) depicts $\mathbb{P}_T(t_d)$ for two characteristic temperatures $\mathcal{T}$, where oscillatory fringes are observed that persist after thermal averaging.
Namely, $\mathbb{P}_T(t_d)$ exhibits fast oscillations throughout regions I and II, and additional slow ones only in region II. 
The contributing frequencies are identified in the Fourier spectra of RTA demonstrated in panels (d) and (e) for regions I and II, respectively.
In region I, independently of the temperature, a single frequency dominates in $\mathbb{P}_T(t_d)$ at
$\omega/(2\pi)=71.8 \, \rm{kHz}$
[\cref{Fig:Spectrum} (d)] corresponding to the energy difference $|E_A^{(1)}-E_T^{(2)}|/h$.
For longer dark times (region II), three distinct frequencies occur, \cref{Fig:Spectrum}(e), with the high ones, i.e. $\omega/(2\pi)=63.7$ and $69.9 \, \rm{kHz}$, referring to the superposition of the first trap state with the first atom-dimer and excited Efimov states, respectively.
The low-frequency peak at $\omega/(2\pi)=6.5 \, \rm{kHz}$ originates from interfering amplitudes between the first atom-dimer and first excited Efimov state pathways.
Note that region II ($\sim 1.2 \, \rm{kHz}$) shows better frequency resolution than region I ($\sim 10 \, \rm{kHz}$), which results in small deviations between the highest frequencies in both regions. Due to the finite resolution, a small mismatch also occurs between the difference $69.9-63.7 \, \rm{kHz}$ and the low frequency peak in region II.
Similar low-frequency and temperature independent oscillatory fringes were also experimentally observed for $^7$Li atoms \cite{yudkin_coherent_2019,yudkin_efimov_2020}.
However, the present analysis reveals that \textit{high-frequency interferences} are also imprinted in the RTA probability, where the \textit{early dark time fringes} can be experimentally utilized to measure the Efimov binding energy at a given $a_{bg}$.

The fact that $\mathbb{P}_T(t_d)$ features three main frequencies, irrespectively of $\mathcal{T}$, is traced back to the incoherent sum of the trimer probability [see \cref{Eq:Efimov_fraction}].  Namely, all  contributions involving higher-lying trap states peter out, except for three arising from the ground trap state $E_A^{(1)}$, the first atom-dimer $E_{AD}^{(1)}$ and the first excited Efimov state $E_T^{(2)}$.
This particular set of eigenstates survives upon the thermal average due to the specifics of the pulse and its envelope.
Recall that the driving frequency is in resonance between the $E_A^{(1)}$ and $E_{AD}^{(1)}$ stationary eigenstates, whereas the duration of the pulse is short in order to coherently populate only the first atom-dimer and first excited Efimov states.

Focusing on this aspect, a three-level model (TLM) Hamiltonian containing $E^{(2)}_{T}$, $E^{(1)}_{AD}$ and a single trap state is constructed \cite{lambropoulos2007fundamentals}.
The three-level system is initialized in the single trap state and we apply square pulses of the scattering length [Eqs. \eqref{Eq:Pulse} \eqref{Eq:Pulse_envelope}] to trigger the dynamics of the three-body setup. Within this picture, the probability amplitude to occupy the first excited Efimov state at the end of the second pulse is obtained by employing first-order time-dependent perturbation theory [for additional details see also \cref{Ap:Three_level}]. Moreover, approximations for the energy levels of the trap states and the matrix elements to occupy the trimer state lead to analytical expressions for $\mathbb{P}_T(t_d)$ [see also \cref{Ap:Matrix_elements} and \ref{Ap:Three_level}]. It is shown that the latter is decomposed into three oscillatory terms.
The TLM predictions for the frequencies, illustrated as vertical dotted lines in Figs. \ref{Fig:Spectrum} (d), (e), are found to be in excellent agreement with the full numerical calculations. 

\begin{figure}[t!]
\centering
\includegraphics[width=0.48 \textwidth]{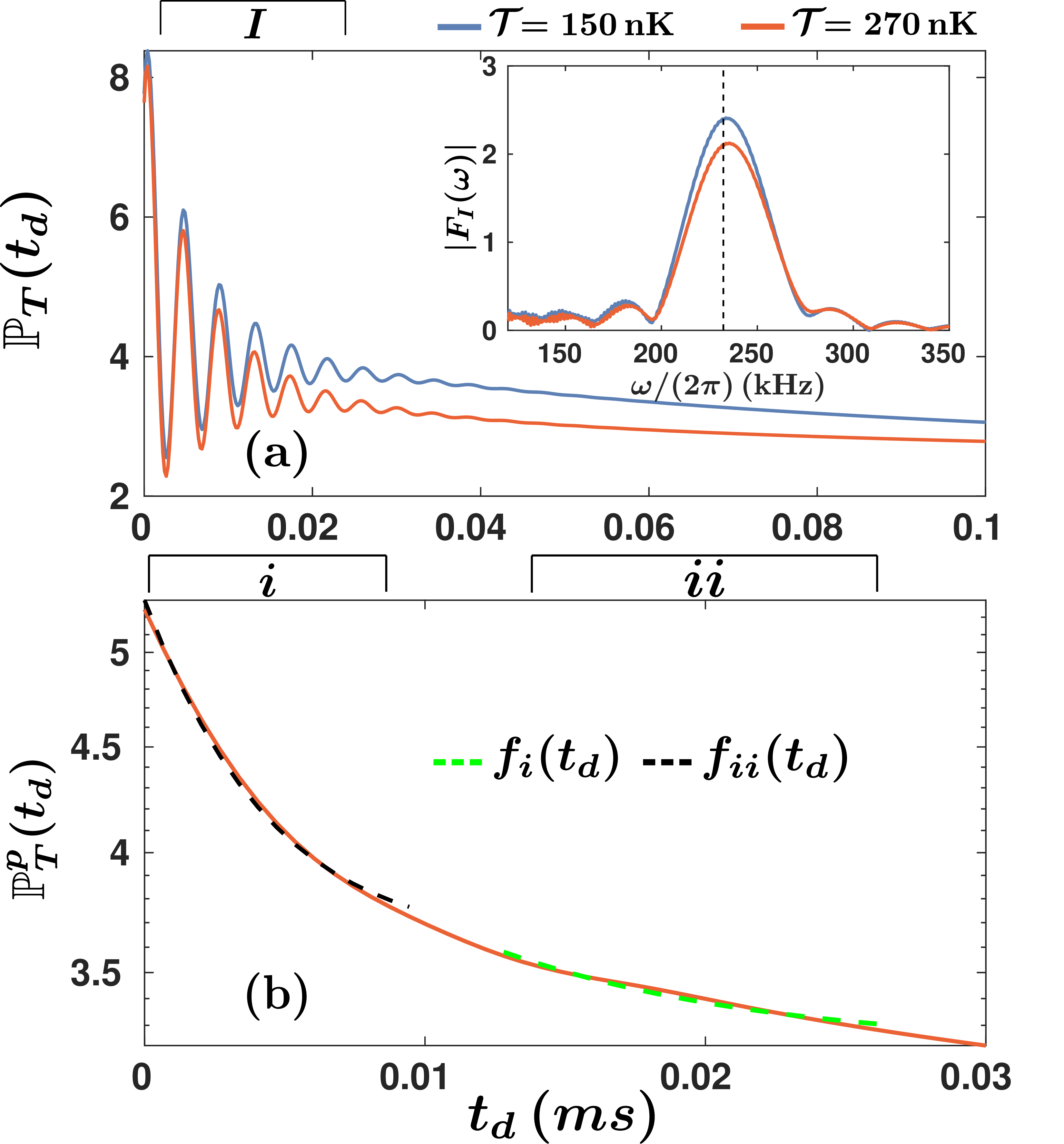}
\caption{(a) $\mathbb{P}_{T}(t_d)$ at $a_{bg}=-2030 \, a_0$ and various temperatures (see legend). 
The driving frequency is resonant with the transition between the ground Efimov and the first trap state, and the decay width of the former $\Gamma^{(1)}/h=41 \, \rm{kHz}$.
The inset presents the frequency spectrum of region I, $\abs{F_I(\omega)}$. 
(b) The mean peak-to-peak envelope, $\mathbb{P}^p _{T}(t_d)$ at $\mathcal{T}=270 \, \rm{nK}$ is fitted with $f_{i/ii}(t_d)=g_{i/ii}e^{-\Gamma_{i/ii}(t_d-t^0_{i/ii})/\hbar}+w_{i/ii}$ at the dark time intervals $i$ and $ii$. 
Even at attractive interactions the energy and lifetime of Efimov states can be simultaneously assessed.}
\label{Fig:Decay_abg_m2030}
\end{figure}

\section{Impact of the lifetime of the trimer}  \label{Sec:Lifetime}

In \cref{Fig:Spectrum}(c)-(e), our analysis neglects the decay of the Efimov trimers and dimer states. 
However, in thermal gases three-body recombination or relaxation processes are present resulting in finite lifetimes of the trimers and dimers.
In the following, we choose $a_{bg}=2030 \, a_0$ that is significantly larger than the van der Waals length scale $l_{VdW}=82.5 \, a_0$ for $^{85}$Rb, yielding negligible finite range effects \cite{chin_feshbach_2010}.
Therefore, in this universal regime, the zero-range theory predicts that the lifetime of the first excited Efimov state is $\hbar/\Gamma^{(2)}=212 \, \mu s$ ($\Gamma^{(2)}$ denotes the decay width) \cite{werner_unitary_2006,petrov_weakly_2004,nielsen_efimov_2002,klauss_observation_2017}.
Also, since the decay of dimers lie within the range  2-9~$ms$, for local peak density $n_0=5 \cdot 10^{12} \, \rm{cm}^{-3}$ \cite{Braaten_enhanced_2004,Claussen_precision_2003,Kohler_spontaneous_2005}, they can be safely neglected within the considered range, $t_d\le 1~\rm{ms}$, rendering the lifetime of Efimov trimers the most relevant decay mechanism.
Furthermore, the pulse frequency is $\Omega/2\pi=10.8 ~\rm{kHz}$ over a time span $2t_0+\tau=134.7 \, \mu s$ ensuring that the Efimov trimers do not decay during the pulse.
Under these considerations, it suffices after the first pulse to multiply the amplitude of the $E_T^{(2)}$ state with the factor $e^{-\Gamma^{(2)}t_d/(2\hbar)}$, as was employed in Refs. \cite{colussi_dynamics_2018,colussi_bunching_2019}.

The interference fringes of the RTA probability including the effect of the decay at 150 and 270 nK are provided in \cref{Fig:Decay_abg_2030a0}(a).
Owing to the large $a_{bg}$, the frequencies are in the range of tenths of kHz adequately agreeing with the TLM calculations [see dashed lines in the inset  \cref{Fig:Decay_abg_2030a0} (a)].
Isolating the impact of the Efimov states decay on the RTA probability, \cref{Fig:Decay_abg_2030a0} (b) shows the mean peak-to-peak envelopes of $\mathbb{P}_T(t_d)$, i.e. $\mathbb{P}^p_{T}(t_d)$.
Fitting $\mathbb{P}^p_{T}(t_d)$ with $f_{i/ii}(t_d)=g_{i/ii}e^{-\Gamma_{i/ii}(t_d-t^0_{i/ii})/\hbar}+w_{i/ii}$ at the dark time intervals $i$ and $ii$ [see dashed lines in \cref{Fig:Decay_abg_2030a0} (b)] reveals two distinct decay widths independent of the temperature.
Namely, $\Gamma_i/h=749.925(1.47) \, \rm{Hz}$ close to $\Gamma^{(2)}/h$, while at later $t_d$, $\Gamma_{ii}/h=375.03(1.63) ~\rm{Hz}$, approximately $\Gamma^{(2)}/(2h)$. 
This means that at early dark times $\mathbb{P}_T(t_d)$ falls off according to the intrinsic lifetime of the $E^{(2)}_T$ Efimov trimer.
In region II, where the interference between the first atom-dimer and the first excited trimer is pronounced, the decay of the RTA probability is \textit{nearly twice} the lifetime of the $E^{(2)}_T$ state.
This effect can in principle explain the unusually long decay times observed in the experiment \cite{yudkin_coherent_2019}. 

Including the trimer's lifetime in the TLM allows to gain insights on the decay of the RTA probability, where $\mathbb{P}_{T}(t_d)$ becomes proportional to
\begin{gather}
\mathbb{P}_{T}(t_d)  \propto [\mathbb{B}_{T,A}(t_d)+\mathbb{B}_{T,AD}(t_d)]e^{-\frac{\Gamma^{(2)}t_d}{2\hbar}}+\mathbb{B}_{AD,A}(t_d)\nonumber \\
+ e^{-\frac{\Gamma^{(2)}t_d}{\hbar}}.
\label{Eq:Signal_decay}
\end{gather}
The terms $\mathbb{B}_{i,j}(t_d)=A_{i,j}(t_d)\sin \left[(E^{(\sigma)}_i-E^{(1)}_j)t_d/\hbar  \right]$ with $\sigma=1+\delta_{i,T}$ originate from the superposition of states $i$, $j$, and $A_{i,j}(t_d)$ refer to  their amplitudes (see details in \cref{Ap:Three_level}). 
The first three terms correspond to the three dominant frequencies shown as dashed lines in the inset of \cref{Fig:Decay_abg_2030a0} (a).
The mixed contributions that involve $E^{(2)}_{T}$ with another state, contain only the factor $e^{-\Gamma^{(2)}t_d/(2\hbar)}$.
Therefore, within region II where the coherent admixture between the $E_{AD}^{(1)}$ and $E_T^{(2)}$ states is manifested, the decay time of $\mathbb{P}_T(t_d)$ is virtually twice as long as the intrinsic Efimov lifetime.
The last non-oscillatory term in \cref{Eq:Signal_decay} involves only the Efimov state and thus decays according to $e^{-\Gamma^{(2)}t_d/\hbar}$. 
The above expression \textit{holds in general} for any atomic species and $a_{bg}>0$, provided that both the first excited Efimov and first atom-dimer are coherently populated. 

As a generalization, the RTA probability is demonstrated in \cref{Fig:Decay_abg_m2030} at negative scattering lengths, e.g. $a_{bg}=-2030 \, a_0$, where the atom-dimer pathways are intrinsically absent since no universal dimer exists.
The pulse frequency $\Omega/2\pi=|E_T^{(1)}-E_A^{(1)}|/h=232.2 ~\rm{kHz}$ and its duration is $2t_0+\tau=3.7 \, \mu s$.
Note that here the pulse resonantly couples the first trap and the Efimov {\it ground} state, whereas the pulse's length is shorter than the ground Efimov state lifetime $\hbar/\Gamma^{(1)}$=3.9 ~$\mu s$ \cite{braaten_universality_2006}. 
As expected, the $\mathbb{P}_{T}(t_d)$ in \cref{Fig:Decay_abg_m2030}(a) oscillates with a single frequency, i.e. $\omega/(2\pi)=\abs{E^{(1)}_T-E^{(1)}_A}/h=233.5 \, \rm{kHz}$, only in region I and vanishes fast due to the large $\Gamma^{(1)}$ decay width. 
Moreover, \cref{Fig:Decay_abg_m2030}(b) showcases the mean peak-to-peak amplitude $\mathbb{P}^p_{T}(t_d)$ and their fittings at the dark time intervals $i$ and $ii$ [see dashed lines in \cref{Fig:Decay_abg_m2030} (b)]. 
Similar to \cref{Fig:Decay_abg_2030a0} (b), we extract two decay widths with their values being $\Gamma_i/h=41.35(5.35) \, \rm{kHz}$ and $\Gamma_{ii}/h=17.56(7.02) \, \rm{kHz}$ at $\mathcal{T}=270 \, \rm{nK}$, which within error bars are close to $\Gamma^{(1)}/h$ and $\Gamma^{(1)}/(2h)$, respectively.
These findings are in accordance to the description of  \cref{Eq:Signal_decay}, omitting terms associated with atom-dimer transitions.
 
 \section{Conclusions and outlook}   \label{Sec:Conclusions}

In summary, the present theory demonstrates that the double magnetic field interferometer has broad applicability. 
Namely, it permits the simultaneous extraction of the binding energy and the lifetime of Efimov states regardless the sign/magnitude of the scattering length and the temperature of the gas.
This is feasible due to the generated superpositions of the trimer with the first atom-dimer and trap state at repulsive interactions, or only with the first trap eigenstate at attractive interactions. These superpositions are manifested as interference (Ramsey) fringes in the probability to occupy trimers, observed over a wide range of temperatures. Corroborating our results, a three-level model is constructed, taking into account only the contributions stemming from the Efimov trimer, the first atom-dimer and trap state.

Going beyond previous studies, our analysis demonstrates that the Ramsey fringes possess long damping times equal to twice the intrinsic lifetime of Efimov trimers.
This behavior is illustrated at long dark times between the pulses, attributed to the superposition of the trimer with the first atom-dimer state. 
This relation in particular provides also an upper bound to the lifetime of $^7$Li Efimov trimers which has remained unknown to date. 
Furthermore, our work predicts that there are additional interference terms surviving the thermal average at short dark times. Namely, in this regime the system exhibits interference fringes with frequencies that coincide with the binding energy of the Efimov states, whereas the decay of these oscillations is dictated by the lifetime of the trimers.
This demonstrates that it is possible to extract the binding energy of the trimer at this early dark time regime, irrespective of the interaction strength. This extends the current experimental practice, exploring the long dark time region~\cite{yudkin_reshape_2023,yudkin_coherent_2019}.

Owing to the sensitivity of the Ramsey-type dynamical protocol, the corresponding interferometric signals could be further employed for probing Efimov states especially at attractive interactions. At this regime, trimers merge with the three-atom continuum at a scattering length related only to the van der Waals length, the so-called van der Waals universality \cite{Etrych_pinpointing_2023,Xie_observation_2020,Berninger_universality_2011,Johansen_testing_2017,Naidon_microscopic_2014,Wang_origin_2012}. The interferometry scheme can thus be utilized at this regime, providing stringent tests on the universality. 
Furthermore, recent experiments explore the modifications of three-body recombination processes in mixtures of a bosonic thermal gas with a degenerate fermion gas \cite{blochprl2022supress}.
Hence, creation of dynamically coherent superpositions between few-body states can reveal the influence of a dense many-body environment on them.

\begin{acknowledgments}
We are grateful to H.R. Sadeghpour and J. P. D'Incao for fruitful discussions.
G. B. acknowledges financial support by the State Graduate Funding Program Scholarships (Hmb-NFG). 
S.I.M. acknowledges support from the NSF through a grant for ITAMP at Harvard University. The Purdue research has been supported in part by the U.S. National Science Foundation, Grant No. PHY-2207977.
This work has been supported by the Cluster of Excellence `The Hamburg Center for Ultrafast Imaging' of the Deutsche Forschungsgemeinschaft (DFG)-EXC 1074- project ID 194651731.
This research was supported in part by the National Science Foundation under Grants No. NSF PHY-1748958 and PHY-2309135. 
\end{acknowledgments}

\appendix

\section{The three-body time-dependent Schr\"odinger equation in hyperspherical coordinates}  \label{Ap:Three_body}

The time-dependent three-body Hamiltonian is decomposed in the center-of-mass frame and further expressed in hyperspherical coordinates.
An expansion in the field-free eigenstates is subsequently utilized to cast the TDSE in matrix form, tackled with the split operator method.

\subsection{Center-of-mass decomposition}  \label{Ap:Three_body_decomposition}

According to Eq.~(1) in the main text, the three-body Hamiltonian in the laboratory frame reads

\begin{gather}
\mathcal{H}(t)=\sum_{i=1}^{3} \left( \frac{-\hbar^2  \nabla_i^2}{2m}  +\frac{m\omega_r^2}{2} \boldsymbol{r}_i^2 \right)  +\sum_{i<j} \frac{4\pi \hbar^2 a(t)}{m} \delta(\boldsymbol{r}_{ij})\hat{O}_{ij}.
\label{Eq:hamilt_lab_Supp}
\end{gather}

In order to eliminate the three degrees of freedom associated to the center-of-mass Hamiltonian we perform a transformation from the laboratory to the center-of-mass frame. The Hamiltonian splits into a time-independent center-of-mass part and another one describing the relative degrees of freedom, i.e. $\mathcal{H}(t)=\mathcal{H}_{cm}+\mathcal{H}_{rel}(t)$.
Evidently, $\mathcal{H}_{rel}(t)$ encapsulates the relevant three-body dynamics, which in hyperspherical coordinates, \cite{rittenhouse_greens_2010,greene_universal_2017,bougas2021few} takes the following expression

\begin{gather}
\mathcal{H}_{rel}(t) = -\frac{\hbar^2}{2\mu} \frac{1}{R^{5/2}}  
\frac{\partial^2}{\partial R^2} \left( R^{5/2} \cdot    \right) +\frac{15 \hbar^2}{8 \mu R^2} +\frac{\hbar^2 \boldsymbol{\Lambda}^2}{2\mu R^2} \nonumber \\
+\frac{1}{2} \mu \omega^2_r R^2 
+V_{bg}(R;\boldsymbol{\varpi}) +V(R;\boldsymbol{\varpi}) f(t).
\label{Eq:Rel_Hamilt_Hyper}
\end{gather}

In this coordinate system, $R$ describes the overall system size, and the five hyperangles collectively indicated by $\boldsymbol{\varpi}$ address the relative particle positions. $V_{bg}(R;\boldsymbol{\varpi})$ and $V(R;\boldsymbol{\varpi})$ are the contact interaction potentials associated to the background ($a_{bg}$) and amplitude scattering length ($a_m$) respectively, expressed in hyperspherical coordinates. Moreover, we have isolated the time-dependence in the function $f(t)=[a(t)-a_{bg}]/a_m$. $\boldsymbol{\Lambda}^2$ is the grand angular momentum operator describing the total angular momentum of the three atoms \cite{avery_hyperspherical_1989}, and $\mu$ is the three-body reduced mass.

According to \cref{Eq:Rel_Hamilt_Hyper}, $\mathcal{H}_{rel}(t)$ splits into a field-free Hamiltonian that describes three particles interacting with $a_{bg}$ scattering length and a time-dependent part which contains the pulse field, i.e. $\mathcal{H}_{rel}(t)=\mathcal{H}_{bg}+V(R;\boldsymbol{\varpi})f(t)$.
This particular structure of $\mathcal{H}_{rel}(t)$ suggests that the time-dependent three-body wave function pertaining to the Hamiltonian \cref{Eq:hamilt_lab_Supp} can be conveniently expanded on the field-free basis set, $\ket{n}$, a basis such that $\mathcal{H}_{bg}$ is a diagonal matrix.

\subsection{Eigenstates of the background Hamiltonian}  \label{Ap:Three_body_background}

Therefore, in order to obtain the eigenstates $\{  \ket{n} \}$ of $\mathcal{H}_{bg}$, we employ the adiabatic hyperspherical representation \mbox{\cite{greene_universal_2017,nielsen_three-body_2001}}, where the hyperradius $R$ is treated as an adiabatic parameter.
For completeness reasons, a brief description on the calculation of $\ket{n}$ in this formalism is provided below.
Namely, $\mathcal{H}_{bg}$ is recasted as follows:

\begin{gather}
\mathcal{H}_{bg}=-\frac{\hbar^2}{2\mu} \frac{1}{R^{5/2}} 
\frac{\partial^2}{\partial R^2} \left( R^{5/2} \cdot    \right)
\nonumber \\
+ \underbrace{\frac{15 \hbar^2}{8 \mu R^2} +\frac{\hbar^2 \boldsymbol{\Lambda}^2}{2\mu R^2} +\frac{1}{2} \mu \omega^2_r R^2 
	+V_{bg}(R;\boldsymbol{\varpi})}_{\mathcal{H}_{ad}(R;\boldsymbol{\varpi)}},
\label{Eq:Hamilt_split}
\end{gather}
where $\mathcal{H}_{ad}(R;\boldsymbol{\varpi})$ refers to the adiabatic hyperangular Hamiltonian which parametrically depends on the hyperradius $R$.
In addition, the eigenstates $\ket{n}$ are expressed by the ansatz
\begin{equation}
\braket{R, \boldsymbol{\varpi} | n} =R^{-5/2} \sum_{\nu} F_{\nu}^{(n)}(R) \Phi_{\nu}(R;\boldsymbol{\varpi}),
\label{Eq:ketn}
\end{equation}
where $F^{(n)}_{\nu}(R)$ [$\Phi_{\nu}(R;\boldsymbol{\varpi})$] denotes the hyperradial [hyperangular] component of $\ket{n}$.
More specifically, $\Phi_{\nu}(R;\boldsymbol{\varpi})$ are obtained by diagonalizing $\mathcal{H}_{ad}(R;\boldsymbol{\varpi})$ at fixed hyperradius $R$ \mbox{\cite{rittenhouse_greens_2010,bougas2021few}} according to the expression
\begin{equation}
\mathcal{H}_{ad}(R;\boldsymbol{\varpi})\Phi_{\nu}(R;\boldsymbol{\varpi})=U_{\nu}(R)\Phi_{\nu}(R;\boldsymbol{\varpi}), 
\label{Eq:Hamilt_ad}
\end{equation}
where $U_{\nu}(R)$ represents the $\nu$-th hyperspherical potential curve that depends only on $R$.
The hyperradial functions $F^{(n)}_{\nu}(R)$ are determined by acting with $ \mathcal{H}_{bg}$ on $\ket{n}$ and integrating over all the hyperangles $\boldsymbol{\varpi}$.
This yields a system of coupled hyperradial equations that include the non-adiabatic couplings \mbox{\cite{rittenhouse_greens_2010,greene_universal_2017}}. By diagonalizing the resulting matrix equations we obtain the eigenenergies $E^{(n)}$ and hyperradial wave functions $F^{(n)}_{\nu}(R)$ \mbox{\cite{rittenhouse_greens_2010,greene_universal_2017}}.

\subsection{Solution of the TDSE}  \label{Ap:Three_body_propagation}

Expanding the time-dependent three-body wave function in terms of $\ket{n}$ yields the following relation:

\begin{equation}
\ket{\Psi^{(\alpha)}_{3b}(t)}=\sum_n c^{(\alpha)}_n(t) \ket{n} \ket{0}_{cm},
\label{Eq:Wavefunction_expansion}
\end{equation}
where the time-dependent coefficients initially satisfy $c^{(\alpha)}_n(t=0)= \delta_{n \alpha}$, and the $\alpha$ index refers to an initial trap state. $\ket{0}_{cm}$ is the center-of-mass ground state.

Plugging \mbox{\cref{Eq:Wavefunction_expansion}} into the TDSE  under the Hamiltonian of  \mbox{\cref{Eq:hamilt_lab_Supp}} leads to a matrix differential equation for the time-dependent expansion coefficients,

\begin{equation}
i\hbar \frac{d \boldsymbol{c}^{(\alpha)}(t)}{dt} = (\boldsymbol{\mathcal{H}_{bg}}+f(t) \boldsymbol{V}) \cdot \boldsymbol{c}^{(\alpha)}(t).
\label{Eq:Differential_equation}
\end{equation}

\cref{Eq:Differential_equation} is solved numerically by utilizing the second-order split-operator method \cite{Burstein_third_1970}.
Namely, the propagator of the $\boldsymbol{c}^{(\alpha)}(t)$ vectors within the time interval ($t$, $t+dt$) reads 
\begin{gather}
\boldsymbol{c}^{(\alpha)}(t+dt)=e^{-i \boldsymbol{\mathcal{H}_{bg}}dt/(2\hbar)} 
e^{-i \boldsymbol{V}/\hbar \int_t^{t+dt} dt' \,  f(t')} \nonumber \\
\times e^{-i \boldsymbol{\mathcal{H}_{bg}}dt/(2\hbar)}  \boldsymbol{c}^{(\alpha)}(t) +\mathcal{O}(dt^3).
\label{Eq:Split_step}
\end{gather}

\section{Matrix elements of the interaction potential with the field-free eigenstates}  \label{Ap:Matrix_elements}

Having at hand the set of field-free eigenstates $\{ \ket{n} \}$, obtained from the adiabatic hyperspherical formalism, the matrix elements of the interaction potential associated to $a_m$, $\boldsymbol{V}_{n' n}$, can be evaluated as 
\begin{gather}
\boldsymbol{V}_{n' n}=\sum_{\nu,\nu'} \int dR \: F^{(n')*}_{\nu'}(R) \mathcal{M}_{\nu' \nu}(R) F^{(n)}_{\nu}(R), \\
\mathcal{M}_{\nu' \nu}(R) =  \braket{\Phi_{\nu'}(R)|V| \Phi_{\nu}(R)}_{\boldsymbol{\varpi}},
\label{Eq:Matrix_elements}
\end{gather}
where $\braket{\cdot}_{\boldsymbol{\varpi}}$ indicates that the integral is performed over the hyperangles.

Eq. \eqref{Eq:Matrix_elements} can be recasted in a simple form by exploiting the property 
$V(R;\boldsymbol{\varpi})=-(a_m/a_{bg})\frac{R}{3} \partial_R V_{bg}(R;\boldsymbol{\varpi})$
between the contact potentials and utilizing the Hellman-Feynman theorem \cite{Feynman_forces_1939}.
Namely, for $\nu\ne\nu'$ the relation 
$\mathcal{M}_{\nu' \nu}(R) =  -(a_m/a_{bg})R\braket{\Phi_{\nu'}(R)|\partial_R\Phi_{\nu}(R)}_{\boldsymbol{\varpi}} [U_{\nu'}(R)-U_\nu(R)]$
holds.
Similar expressions are derived for $\nu=\nu'$ which can be regrouped as follows

\begin{equation}
\mathcal{M}_{\nu' \nu}(R)=\frac{a_m}{a_{bg}} \frac{\hbar^2}{2\mu R} (-)^{1+\rm{sgn}(\nu-\nu')} \sqrt{\partial_R s^2_{\nu}(R) \partial_R s^2_{\nu'}(R)}.
\label{Eq:Hellman_Feynman}
\end{equation}
Here, $s^2_{\nu}(R)$ are related to the potential curves, i.e. $2\mu R^2 / \hbar^2 U_{\nu} (R) = s^2_{\nu}(R)-1/4$, and $\rm{sgn}(\cdot)$ denotes the sign function. 

\section{Three-level model and perturbation theory} \label{Ap:Three_level}

To provide a simplified picture of the full dynamics of the few-body bound states we next construct an effective three-level model.
Within this model, 
we consider only three field-free eigenstates, the first excited Efimov trimer (T), the first atom-dimer (AD) and an initial trap state $\alpha$.

At the end of the first pulse, the probability amplitude to occupy the $T$ state, $\Bar{c}^{(\alpha)}_T$, within first-order time-dependent perturbation theory~\cite{sakurai1967advanced}, reads

\begin{subequations}
	\begin{gather}
	\Bar{c}^{(\alpha)}_T(t_0+\tau)= \boldsymbol{V}_{T,\alpha} R_{T,\alpha}(t_0+\tau),  \\
	\label{Eq:First_pulse}
	R_{n,m}(t_0+\tau) = \frac{-e^{i(\omega_{n,m}+\Omega)(t_0+\tau)/2}\sin\left[ (\omega_{n,m}+\Omega)\frac{t_0+\tau}{2} \right]}{\hbar(\omega_{n,m}+\Omega)} \nonumber \\
	-(\Omega \leftrightarrow - \Omega), 
	\end{gather}
\end{subequations}
where $\omega_{n,m} \equiv (E^{(n)}-E^{(m)})/\hbar$. 

During the dark time $t_d$, the probability amplitude of the $n$-th state acquires the phase factor 
$e^{-i E^{(n)}t_d/ \hbar} \Bar{c}^{(\alpha)}_n(t_0+\tau) $. 
In particular, the amplitude of the first excited Efimov state is supplemented with the factor $e^{-\Gamma^{(2)}t_d/(2\hbar)}$, due to the width $\Gamma^{(2)}$ of the Efimov state, leading to the decay of the latter during $t_d$.

The second pulse mixes all states together, and the probability amplitude to occupy the $T$ state at the end of this pulse reads,
\begin{gather}
\Bar{d}^{(\alpha)}_T(2t_0+2\tau+t_d) = \sum_{j=T,AD} \Big[ \boldsymbol{V}_{T,j} R_{T,j}(t_0+\tau) \nonumber \\
\times  \Bar{c}^{(\alpha)}_j(t_0+\tau) e^{-iE^{(\sigma)}_j t_d/\hbar-\Gamma^{(2)}t_d/(2\hbar) \delta_{T,j}} \Big] \nonumber \\
+\boldsymbol{V}_{T,\alpha} R_{T,\alpha}(t_0+\tau) \Bar{c}^{(\alpha)}_A(t_0+\tau) e^{-i E^{(\alpha)}_A t_d/\hbar}, 
\label{Eq:Second_pulse}
\end{gather}
where $\sigma=1+\delta_{j,T}$.

To obtain the ratio of the thermally averaged probability $\mathbb{P}_T(t_d)$, we weight the probabilities $\abs{\Bar{d}^{(\alpha)}_T(2t_0+2\tau+t_d)}^2$ and $\abs{\Bar{c}^{(\alpha)}_T (t_0+\tau)}^2$ according to the Maxwell-Boltzman distribution for the trap states of energy $E_A^{(\alpha)}$ at temperature $\mathcal{T}$,
\begin{equation}
\mathbb{P}_T(t_d) = \frac{  \sum_{\alpha \in A}  e^{-\frac{E_A^{(\alpha)}}{k_B  \mathcal{T}}}  \abs{\Bar{d}^{(\alpha)}_T(2t_0+2\tau+t_d)}^2  }{   \sum_{\alpha \in A}  e^{-\frac{E_A^{(\alpha)}}{k_B  \mathcal{T}}} \abs{\Bar{c}^{(\alpha)}_T (t_0+\tau)}^2 },
\label{Eq:Thermal_average_expression}
\end{equation}
where $k_B$ is the Boltzmann constant. 

In order to derive an analytical expression for \cref{Eq:Thermal_average_expression} additional approximations are used.
Namely, the expressions for $\Bar{d}_T^{(\alpha)}(2t_0+2\tau+t_d)$ and $\Bar{c}^{(\alpha)}_T(t_0+\tau)$ can be further simplified by employing the rotating-wave approximation \cite{sakurai1967advanced}.

Furthermore, the energy of the $\alpha$-th trap state is roughly approximated by the non-interacting energy spectrum, $E_A^{(\alpha)}=E_A^{(1)}+2\alpha \hbar \omega_{r}$,
where $E^{(1)}_{A}$ is the energy of the first trap state. 
In addition, we approximate the $\boldsymbol{V}_{T,\alpha}$ matrix elements with a quartic root of the energy of the $\alpha$-th trap state, a dependence corroborated by a fitting procedure.
Under these considerations, Eq. \eqref{Eq:Thermal_average_expression} obtains the same form as Eq. (5) in the main text,

\begin{widetext}
	
	\begin{equation}
	\mathbb{P}_T(t_d) \propto  [\mathbb{B}_{T,A}(t_d)+ \mathbb{B}_{T,AD}(t_d)]e^{-\Gamma^{(2)}t_d/(2\hbar)} +\mathbb{B}_{AD,A}(t_d) 
	+   e^{-\Gamma^{(2)} t_d/\hbar}, 
	\label{Eq:Decomposition} 
	\end{equation}
	where the $\mathbb{B}$-terms are given by the expressions
	
	\begin{subequations}
		\begin{gather}
		\mathbb{B}_{T,A}(t_d)  = C_1 \: \Im \Bigg[ e^{-i\Delta \phi_1} \Phi\left( e^{f(k_B \mathcal{T},t_d,\omega_{r})},-0.5,\frac{E^{(1)}_{A}}{2\hbar \omega_{r}} \right) \Bigg]  \label{Eq:Fast_component_1} \\
		\mathbb{B}_{T,AD}(t_d) = \sum_{\pm} (-)^{\pm} C_2^{\pm} \sin\left[ (E^{(2)}_{T}-E^{(1)}_{AD})t_d/\hbar \pm\Omega(t_0+\tau)/2 \right]    \label{Eq:Slow_component} \\
		\mathbb{B}_{AD,A}(t_d)  = \sum_{\pm}C_3^{\pm} \, \Re \Bigg[  e^{-i\Delta \phi_2 \pm i\Omega(t_0+\tau)/2} 
		\Phi \left(  e^{f(k_B \mathcal{T},t_d,\omega_{r})},-0.5,\frac{E^{(1)}_{A}}{2\hbar \omega_{r}} \right)  \Bigg] \label{Eq:Fast_component_2}  \\
		f(k_B \mathcal{T},t_d,\omega_{r}) = -\frac{2\hbar \omega_{r}}{k_B \mathcal{T}} +2i\omega_{r} [t_d+1.5(t_0+\tau)].
		\end{gather}
	\end{subequations}
\end{widetext}

$\Phi(a,b,z)$ is the Hurwitz-Lersch zeta function \cite{Gradshteyn_table_2015} and the phases $\Delta \phi_1$ and $\Delta \phi_2$ are defined as follows,
\begin{gather}
\Delta \phi_1 \equiv \frac{(E^{(2)}_{T}-E^{(1)}_A)t_d}{\hbar} -3E^{(1)}_A\frac{t_0+\tau}{2\hbar} \\
\Delta \phi_2 \equiv \frac{(E^{(1)}_{AD}-E^{(1)}_A)t_d}{\hbar} -3E^{(1)}_A\frac{t_0+\tau}{2\hbar}. 
\label{Eq:Phase_definitions}
\end{gather}

The explicit form of the prefactors $C_1, C_2^{\pm},C_3^{\pm}$ is given by,

\begin{gather}
C_1=\frac{\hbar \Omega}{\boldsymbol{V}_{T,T}} 
\frac{\left[\Phi \left( e^{-2\hbar \omega_{r}/(k_B \mathcal{T})},-0.5,\frac{E^{(1)}_{A}}{2\hbar \omega_{r}}  \right) \right]^{-1}}{\sin^2[\Omega(t_0+\tau)/2]},
\label{Eq:Prefactor_A}
\end{gather}

\begin{gather}
C_2^{\pm} = \frac{\boldsymbol{V}_{T,AD}}{\boldsymbol{V}_{T,T}}
\frac{\hbar \Omega}{\sin^2[\Omega (t_0+\tau)/2]} \nonumber \\ \times \frac{\sin[(\omega_{T,AD} \pm \Omega)(t_0+\tau)/2]}{\hbar(\omega_{T,AD} \pm \Omega)},
\label{Eq:Prefactor_C}
\end{gather}

\begin{gather}
C_3^{\pm} = (-)^{\pm} \frac{\boldsymbol{V}_{T,AD}}{\abs{\boldsymbol{V}_{T,T}}^2}
\frac{\sin[(\omega_{T,AD} \pm \Omega)(t_0+\tau)/2]}{2\hbar(\omega_{T,AD} \pm \Omega)} \nonumber \\
\frac{\hbar^2 \Omega^2}{\sin^4[\Omega(t_0+\tau)/2]}
\times \left[ \Phi \left( e^{-2\hbar \omega_{r}/(k_B \mathcal{T})},-0.5,\frac{E^{(1)}_{A}}{2\hbar \omega_{r}}  \right) \right]^{-1},
\label{Eq:Prefactor_B}
\end{gather}

Note that there are revivals of the oscillatory signals $\mathbb{B}_{T,A}(t_d)$ and $\mathbb{B}_{AD,A}(t_d)$ at later dark times $\frac{n \pi}{\omega_{r}}-1.5(t_0+\tau)$, which are attributed to the trap \cite{dincao_efimov_2018}.

\bibliography{Three_Bodies_dynamics.bib}

\begin{thebibliography}{60}%
\makeatletter
\providecommand \@ifxundefined [1]{%
 \@ifx{#1\undefined}
}%
\providecommand \@ifnum [1]{%
 \ifnum #1\expandafter \@firstoftwo
 \else \expandafter \@secondoftwo
 \fi
}%
\providecommand \@ifx [1]{%
 \ifx #1\expandafter \@firstoftwo
 \else \expandafter \@secondoftwo
 \fi
}%
\providecommand \natexlab [1]{#1}%
\providecommand \enquote  [1]{``#1''}%
\providecommand \bibnamefont  [1]{#1}%
\providecommand \bibfnamefont [1]{#1}%
\providecommand \citenamefont [1]{#1}%
\providecommand \href@noop [0]{\@secondoftwo}%
\providecommand \href [0]{\begingroup \@sanitize@url \@href}%
\providecommand \@href[1]{\@@startlink{#1}\@@href}%
\providecommand \@@href[1]{\endgroup#1\@@endlink}%
\providecommand \@sanitize@url [0]{\catcode `\\12\catcode `\$12\catcode
  `\&12\catcode `\#12\catcode `\^12\catcode `\_12\catcode `\%12\relax}%
\providecommand \@@startlink[1]{}%
\providecommand \@@endlink[0]{}%
\providecommand \url  [0]{\begingroup\@sanitize@url \@url }%
\providecommand \@url [1]{\endgroup\@href {#1}{\urlprefix }}%
\providecommand \urlprefix  [0]{URL }%
\providecommand \Eprint [0]{\href }%
\providecommand \doibase [0]{http://dx.doi.org/}%
\providecommand \selectlanguage [0]{\@gobble}%
\providecommand \bibinfo  [0]{\@secondoftwo}%
\providecommand \bibfield  [0]{\@secondoftwo}%
\providecommand \translation [1]{[#1]}%
\providecommand \BibitemOpen [0]{}%
\providecommand \bibitemStop [0]{}%
\providecommand \bibitemNoStop [0]{.\EOS\space}%
\providecommand \EOS [0]{\spacefactor3000\relax}%
\providecommand \BibitemShut  [1]{\csname bibitem#1\endcsname}%
\let\auto@bib@innerbib\@empty
\bibitem [{\citenamefont {Efimov}(1970)}]{Efimov_energy_1970}%
  \BibitemOpen
  \bibfield  {author} {\bibinfo {author} {\bibfnamefont {V.}~\bibnamefont
  {Efimov}},\ }\bibfield  {title} {\enquote {\bibinfo {title} {Energy levels
  arising from resonant two-body forces in a three-body system},}\ }\href
  {\doibase https://doi.org/10.1016/0370-2693(70)90349-7} {\bibfield  {journal}
  {\bibinfo  {journal} {Phys. Lett. B}\ }\textbf {\bibinfo {volume} {33}},\
  \bibinfo {pages} {563--564} (\bibinfo {year} {1970})}\BibitemShut {NoStop}%
\bibitem [{\citenamefont {Efimov}(1973)}]{Efimov_energy_1973}%
  \BibitemOpen
  \bibfield  {author} {\bibinfo {author} {\bibfnamefont {V.}~\bibnamefont
  {Efimov}},\ }\bibfield  {title} {\enquote {\bibinfo {title} {Energy levels of
  three resonantly interacting particles},}\ }\href@noop {} {\bibfield
  {journal} {\bibinfo  {journal} {Nucl. Phys. A}\ }\textbf {\bibinfo {volume}
  {210}},\ \bibinfo {pages} {157--188} (\bibinfo {year} {1973})}\BibitemShut
  {NoStop}%
\bibitem [{\citenamefont {Efimov}(1971)}]{Efimov_weakly_1971}%
  \BibitemOpen
  \bibfield  {author} {\bibinfo {author} {\bibfnamefont {V.~N.}\ \bibnamefont
  {Efimov}},\ }\bibfield  {title} {\enquote {\bibinfo {title} {Weakly bound
  states of three resonantly interacting particles.}}\ }\href@noop {}
  {\bibfield  {journal} {\bibinfo  {journal} {Sov. J. Nucl. Phys.}\ }\textbf
  {\bibinfo {volume} {12}},\ \bibinfo {pages} {589} (\bibinfo {year}
  {1971})}\BibitemShut {NoStop}%
\bibitem [{\citenamefont {Greene}\ \emph {et~al.}(2017)\citenamefont {Greene},
  \citenamefont {Giannakeas},\ and\ \citenamefont
  {P\'erez-R\'{\i}os}}]{greene_universal_2017}%
  \BibitemOpen
  \bibfield  {author} {\bibinfo {author} {\bibfnamefont {C.~H.}\ \bibnamefont
  {Greene}}, \bibinfo {author} {\bibfnamefont {P.}~\bibnamefont {Giannakeas}},
  \ and\ \bibinfo {author} {\bibfnamefont {J.}~\bibnamefont
  {P\'erez-R\'{\i}os}},\ }\bibfield  {title} {\enquote {\bibinfo {title}
  {Universal few-body physics and cluster formation},}\ }\href {\doibase
  10.1103/RevModPhys.89.035006} {\bibfield  {journal} {\bibinfo  {journal}
  {Rev. Mod. Phys.}\ }\textbf {\bibinfo {volume} {89}},\ \bibinfo {pages}
  {035006} (\bibinfo {year} {2017})}\BibitemShut {NoStop}%
\bibitem [{\citenamefont {Nielsen}\ \emph {et~al.}(2001)\citenamefont
  {Nielsen}, \citenamefont {Fedorov}, \citenamefont {Jensen},\ and\
  \citenamefont {Garrido}}]{nielsen_three-body_2001}%
  \BibitemOpen
  \bibfield  {author} {\bibinfo {author} {\bibfnamefont {E.}~\bibnamefont
  {Nielsen}}, \bibinfo {author} {\bibfnamefont {D.~V.}\ \bibnamefont
  {Fedorov}}, \bibinfo {author} {\bibfnamefont {A.~S.}\ \bibnamefont {Jensen}},
  \ and\ \bibinfo {author} {\bibfnamefont {E.}~\bibnamefont {Garrido}},\
  }\bibfield  {title} {\enquote {\bibinfo {title} {The three-body problem with
  short-range interactions},}\ }\href {\doibase 10.1016/S0370-1573(00)00107-1}
  {\bibfield  {journal} {\bibinfo  {journal} {Phys. Rep.}\ }\textbf {\bibinfo
  {volume} {347}},\ \bibinfo {pages} {373--459} (\bibinfo {year}
  {2001})}\BibitemShut {NoStop}%
\bibitem [{\citenamefont {Naidon}\ and\ \citenamefont
  {Endo}(2017)}]{naidon_efimov_2017}%
  \BibitemOpen
  \bibfield  {author} {\bibinfo {author} {\bibfnamefont {P.}~\bibnamefont
  {Naidon}}\ and\ \bibinfo {author} {\bibfnamefont {S.}~\bibnamefont {Endo}},\
  }\bibfield  {title} {\enquote {\bibinfo {title} {Efimov physics: a review},}\
  }\href {\doibase 10.1088/1361-6633/aa50e8} {\bibfield  {journal} {\bibinfo
  {journal} {Rep. Prog. Phys.}\ }\textbf {\bibinfo {volume} {80}},\ \bibinfo
  {pages} {056001} (\bibinfo {year} {2017})}\BibitemShut {NoStop}%
\bibitem [{\citenamefont {D'Incao}(2018)}]{dincao_few-body_2018}%
  \BibitemOpen
  \bibfield  {author} {\bibinfo {author} {\bibfnamefont {J.~P.}\ \bibnamefont
  {D'Incao}},\ }\bibfield  {title} {\enquote {\bibinfo {title} {Few-body
  physics in resonantly interacting ultracold quantum gases},}\ }\href
  {\doibase 10.1088/1361-6455/aaa116} {\bibfield  {journal} {\bibinfo
  {journal} {J. Phys. B: At. Mol. Opt. Phys.}\ }\textbf {\bibinfo {volume}
  {51}},\ \bibinfo {pages} {043001} (\bibinfo {year} {2018})}\BibitemShut
  {NoStop}%
\bibitem [{\citenamefont {Kraemer}\ \emph {et~al.}(2006)\citenamefont
  {Kraemer}, \citenamefont {Mark}, \citenamefont {Waldburger}, \citenamefont
  {Danzl}, \citenamefont {Chin}, \citenamefont {Engeser}, \citenamefont
  {Lange}, \citenamefont {Pilch}, \citenamefont {Jaakkola}, \citenamefont
  {Nägerl},\ and\ \citenamefont {Grimm}}]{kraemer_evidence_2006}%
  \BibitemOpen
  \bibfield  {author} {\bibinfo {author} {\bibfnamefont {T.}~\bibnamefont
  {Kraemer}}, \bibinfo {author} {\bibfnamefont {M.}~\bibnamefont {Mark}},
  \bibinfo {author} {\bibfnamefont {P.}~\bibnamefont {Waldburger}}, \bibinfo
  {author} {\bibfnamefont {J.~G.}\ \bibnamefont {Danzl}}, \bibinfo {author}
  {\bibfnamefont {C.}~\bibnamefont {Chin}}, \bibinfo {author} {\bibfnamefont
  {B.}~\bibnamefont {Engeser}}, \bibinfo {author} {\bibfnamefont {A.~D.}\
  \bibnamefont {Lange}}, \bibinfo {author} {\bibfnamefont {K.}~\bibnamefont
  {Pilch}}, \bibinfo {author} {\bibfnamefont {A.}~\bibnamefont {Jaakkola}},
  \bibinfo {author} {\bibfnamefont {H.-C.}\ \bibnamefont {Nägerl}}, \ and\
  \bibinfo {author} {\bibfnamefont {R.}~\bibnamefont {Grimm}},\ }\bibfield
  {title} {\enquote {\bibinfo {title} {Evidence for efimov quantum states in an
  ultracold gas of caesium atoms},}\ }\href {\doibase 10.1038/nature04626}
  {\bibfield  {journal} {\bibinfo  {journal} {Nature}\ }\textbf {\bibinfo
  {volume} {440}},\ \bibinfo {pages} {315--318} (\bibinfo {year}
  {2006})}\BibitemShut {NoStop}%
\bibitem [{\citenamefont {Kunitski}\ \emph {et~al.}(2015)\citenamefont
  {Kunitski}, \citenamefont {Zeller}, \citenamefont {Voigtsberger},
  \citenamefont {Kalinin}, \citenamefont {Schmidt}, \citenamefont {Schöffler},
  \citenamefont {Czasch}, \citenamefont {Schöllkopf}, \citenamefont
  {Grisenti}, \citenamefont {Jahnke}, \citenamefont {Blume},\ and\
  \citenamefont {Dörner}}]{Kunitski_observation_2015}%
  \BibitemOpen
  \bibfield  {author} {\bibinfo {author} {\bibfnamefont {M.}~\bibnamefont
  {Kunitski}}, \bibinfo {author} {\bibfnamefont {S.}~\bibnamefont {Zeller}},
  \bibinfo {author} {\bibfnamefont {J.}~\bibnamefont {Voigtsberger}}, \bibinfo
  {author} {\bibfnamefont {A.}~\bibnamefont {Kalinin}}, \bibinfo {author}
  {\bibfnamefont {L.~Ph.~H.}\ \bibnamefont {Schmidt}}, \bibinfo {author}
  {\bibfnamefont {M.}~\bibnamefont {Schöffler}}, \bibinfo {author}
  {\bibfnamefont {A.}~\bibnamefont {Czasch}}, \bibinfo {author} {\bibfnamefont
  {W.}~\bibnamefont {Schöllkopf}}, \bibinfo {author} {\bibfnamefont {R.~E.}\
  \bibnamefont {Grisenti}}, \bibinfo {author} {\bibfnamefont {T.}~\bibnamefont
  {Jahnke}}, \bibinfo {author} {\bibfnamefont {D.}~\bibnamefont {Blume}}, \
  and\ \bibinfo {author} {\bibfnamefont {R.}~\bibnamefont {Dörner}},\
  }\bibfield  {title} {\enquote {\bibinfo {title} {Observation of the efimov
  state of the helium trimer},}\ }\href@noop {} {\bibfield  {journal} {\bibinfo
   {journal} {Science}\ }\textbf {\bibinfo {volume} {348}},\ \bibinfo {pages}
  {551--555} (\bibinfo {year} {2015})}\BibitemShut {NoStop}%
\bibitem [{\citenamefont {Endo}\ \emph {et~al.}(2016)\citenamefont {Endo},
  \citenamefont {Garc\'{\i}a-Garc\'{\i}a},\ and\ \citenamefont
  {Naidon}}]{Endo_universal_2016}%
  \BibitemOpen
  \bibfield  {author} {\bibinfo {author} {\bibfnamefont {S.}~\bibnamefont
  {Endo}}, \bibinfo {author} {\bibfnamefont {A.~M.}\ \bibnamefont
  {Garc\'{\i}a-Garc\'{\i}a}}, \ and\ \bibinfo {author} {\bibfnamefont
  {P.}~\bibnamefont {Naidon}},\ }\bibfield  {title} {\enquote {\bibinfo {title}
  {Universal clusters as building blocks of stable quantum matter},}\
  }\href@noop {} {\bibfield  {journal} {\bibinfo  {journal} {Phys. Rev. A}\
  }\textbf {\bibinfo {volume} {93}},\ \bibinfo {pages} {053611} (\bibinfo
  {year} {2016})}\BibitemShut {NoStop}%
\bibitem [{\citenamefont {Kievsky}\ \emph {et~al.}(2021)\citenamefont
  {Kievsky}, \citenamefont {Gattobigio}, \citenamefont {Girlanda},\ and\
  \citenamefont {Viviani}}]{Kievsky_efimov_2021}%
  \BibitemOpen
  \bibfield  {author} {\bibinfo {author} {\bibfnamefont {A.}~\bibnamefont
  {Kievsky}}, \bibinfo {author} {\bibfnamefont {M.}~\bibnamefont {Gattobigio}},
  \bibinfo {author} {\bibfnamefont {L.}~\bibnamefont {Girlanda}}, \ and\
  \bibinfo {author} {\bibfnamefont {M.}~\bibnamefont {Viviani}},\ }\bibfield
  {title} {\enquote {\bibinfo {title} {Efimov physics and connections to
  nuclear physics},}\ }\href {\doibase 10.1146/annurev-nucl-102419-032845}
  {\bibfield  {journal} {\bibinfo  {journal} {Annu. Rev. Nucl. Part. Sci.}\
  }\textbf {\bibinfo {volume} {71}},\ \bibinfo {pages} {465--490} (\bibinfo
  {year} {2021})}\BibitemShut {NoStop}%
\bibitem [{\citenamefont {Nishida}\ \emph {et~al.}(2013)\citenamefont
  {Nishida}, \citenamefont {Kato},\ and\ \citenamefont
  {Batista}}]{nishida2013efimov}%
  \BibitemOpen
  \bibfield  {author} {\bibinfo {author} {\bibfnamefont {Y.}~\bibnamefont
  {Nishida}}, \bibinfo {author} {\bibfnamefont {Y.}~\bibnamefont {Kato}}, \
  and\ \bibinfo {author} {\bibfnamefont {C.~D.}\ \bibnamefont {Batista}},\
  }\bibfield  {title} {\enquote {\bibinfo {title} {Efimov effect in quantum
  magnets},}\ }\href@noop {} {\bibfield  {journal} {\bibinfo  {journal} {Nature
  Phys.}\ }\textbf {\bibinfo {volume} {9}},\ \bibinfo {pages} {93--97}
  (\bibinfo {year} {2013})}\BibitemShut {NoStop}%
\bibitem [{\citenamefont {Gullans}\ \emph {et~al.}(2017)\citenamefont
  {Gullans}, \citenamefont {Diehl}, \citenamefont {Rittenhouse}, \citenamefont
  {Ruzic}, \citenamefont {D’Incao}, \citenamefont {Julienne}, \citenamefont
  {Gorshkov},\ and\ \citenamefont {Taylor}}]{gullans2017efimov}%
  \BibitemOpen
  \bibfield  {author} {\bibinfo {author} {\bibfnamefont {M.J.}\ \bibnamefont
  {Gullans}}, \bibinfo {author} {\bibfnamefont {S.}~\bibnamefont {Diehl}},
  \bibinfo {author} {\bibfnamefont {S.T.}\ \bibnamefont {Rittenhouse}},
  \bibinfo {author} {\bibfnamefont {B.P.}\ \bibnamefont {Ruzic}}, \bibinfo
  {author} {\bibfnamefont {J.P.}\ \bibnamefont {D’Incao}}, \bibinfo {author}
  {\bibfnamefont {P.}~\bibnamefont {Julienne}}, \bibinfo {author}
  {\bibfnamefont {A.V.}\ \bibnamefont {Gorshkov}}, \ and\ \bibinfo {author}
  {\bibfnamefont {J.M.}\ \bibnamefont {Taylor}},\ }\bibfield  {title} {\enquote
  {\bibinfo {title} {Efimov states of strongly interacting photons},}\
  }\href@noop {} {\bibfield  {journal} {\bibinfo  {journal} {Phys. Rev. Lett.}\
  }\textbf {\bibinfo {volume} {119}},\ \bibinfo {pages} {233601} (\bibinfo
  {year} {2017})}\BibitemShut {NoStop}%
\bibitem [{\citenamefont {Tran}\ \emph {et~al.}(2021)\citenamefont {Tran},
  \citenamefont {Rautenberg}, \citenamefont {Gerken}, \citenamefont {Lippi},
  \citenamefont {Zhu}, \citenamefont {Ulmanis}, \citenamefont {Drescher},
  \citenamefont {Salmhofer}, \citenamefont {Enss},\ and\ \citenamefont
  {Weidemüller}}]{tran_fermions_2021}%
  \BibitemOpen
  \bibfield  {author} {\bibinfo {author} {\bibfnamefont {B.}~\bibnamefont
  {Tran}}, \bibinfo {author} {\bibfnamefont {M.}~\bibnamefont {Rautenberg}},
  \bibinfo {author} {\bibfnamefont {M.}~\bibnamefont {Gerken}}, \bibinfo
  {author} {\bibfnamefont {E.}~\bibnamefont {Lippi}}, \bibinfo {author}
  {\bibfnamefont {B.}~\bibnamefont {Zhu}}, \bibinfo {author} {\bibfnamefont
  {J.}~\bibnamefont {Ulmanis}}, \bibinfo {author} {\bibfnamefont
  {M.}~\bibnamefont {Drescher}}, \bibinfo {author} {\bibfnamefont
  {M.}~\bibnamefont {Salmhofer}}, \bibinfo {author} {\bibfnamefont
  {T.}~\bibnamefont {Enss}}, \ and\ \bibinfo {author} {\bibfnamefont
  {M.}~\bibnamefont {Weidemüller}},\ }\bibfield  {title} {\enquote {\bibinfo
  {title} {Fermions meet two bosons—the heteronuclear efimov effect
  revisited},}\ }\href@noop {} {\bibfield  {journal} {\bibinfo  {journal}
  {Braz. J. Phys.}\ }\textbf {\bibinfo {volume} {51}},\ \bibinfo {pages}
  {316--322} (\bibinfo {year} {2021})}\BibitemShut {NoStop}%
\bibitem [{\citenamefont {Christianen}\ \emph {et~al.}(2022)\citenamefont
  {Christianen}, \citenamefont {Cirac},\ and\ \citenamefont
  {Schmidt}}]{Christianen_bose_2022}%
  \BibitemOpen
  \bibfield  {author} {\bibinfo {author} {\bibfnamefont {A.}~\bibnamefont
  {Christianen}}, \bibinfo {author} {\bibfnamefont {J.~I.}\ \bibnamefont
  {Cirac}}, \ and\ \bibinfo {author} {\bibfnamefont {R.}~\bibnamefont
  {Schmidt}},\ }\bibfield  {title} {\enquote {\bibinfo {title} {Bose polaron
  and the efimov effect: A gaussian-state approach},}\ }\href@noop {}
  {\bibfield  {journal} {\bibinfo  {journal} {Phys. Rev. A}\ }\textbf {\bibinfo
  {volume} {105}},\ \bibinfo {pages} {053302} (\bibinfo {year}
  {2022})}\BibitemShut {NoStop}%
\bibitem [{\citenamefont {Naidon}(2018)}]{Naidon_two_2018}%
  \BibitemOpen
  \bibfield  {author} {\bibinfo {author} {\bibfnamefont {P.}~\bibnamefont
  {Naidon}},\ }\bibfield  {title} {\enquote {\bibinfo {title} {Two impurities
  in a bose--einstein condensate: From yukawa to efimov attracted polarons},}\
  }\href@noop {} {\bibfield  {journal} {\bibinfo  {journal} {J. Phys. Soc.
  Jpn.}\ }\textbf {\bibinfo {volume} {87}},\ \bibinfo {pages} {043002}
  (\bibinfo {year} {2018})}\BibitemShut {NoStop}%
\bibitem [{\citenamefont {Sun}\ and\ \citenamefont
  {Cui}(2019)}]{Sun_efimov_2019}%
  \BibitemOpen
  \bibfield  {author} {\bibinfo {author} {\bibfnamefont {M.}~\bibnamefont
  {Sun}}\ and\ \bibinfo {author} {\bibfnamefont {X.}~\bibnamefont {Cui}},\
  }\bibfield  {title} {\enquote {\bibinfo {title} {Efimov physics in the
  presence of a fermi sea},}\ }\href@noop {} {\bibfield  {journal} {\bibinfo
  {journal} {Phys. Rev. A}\ }\textbf {\bibinfo {volume} {99}},\ \bibinfo
  {pages} {060701} (\bibinfo {year} {2019})}\BibitemShut {NoStop}%
\bibitem [{\citenamefont {Musolino}\ \emph {et~al.}(2022)\citenamefont
  {Musolino}, \citenamefont {Kurkjian}, \citenamefont {Van~Regemortel},
  \citenamefont {Wouters}, \citenamefont {Kokkelmans},\ and\ \citenamefont
  {Colussi}}]{Musolino_bose_2022}%
  \BibitemOpen
  \bibfield  {author} {\bibinfo {author} {\bibfnamefont {S.}~\bibnamefont
  {Musolino}}, \bibinfo {author} {\bibfnamefont {H.}~\bibnamefont {Kurkjian}},
  \bibinfo {author} {\bibfnamefont {M.}~\bibnamefont {Van~Regemortel}},
  \bibinfo {author} {\bibfnamefont {M.}~\bibnamefont {Wouters}}, \bibinfo
  {author} {\bibfnamefont {S.~J. J. M.~F.}\ \bibnamefont {Kokkelmans}}, \ and\
  \bibinfo {author} {\bibfnamefont {V.~E.}\ \bibnamefont {Colussi}},\
  }\bibfield  {title} {\enquote {\bibinfo {title} {Bose-einstein condensation
  of efimovian triples in the unitary bose gas},}\ }\href@noop {} {\bibfield
  {journal} {\bibinfo  {journal} {Phys. Rev. Lett.}\ }\textbf {\bibinfo
  {volume} {128}},\ \bibinfo {pages} {020401} (\bibinfo {year}
  {2022})}\BibitemShut {NoStop}%
\bibitem [{\citenamefont {Colussi}\ \emph
  {et~al.}(2018{\natexlab{a}})\citenamefont {Colussi}, \citenamefont
  {Musolino},\ and\ \citenamefont {Kokkelmans}}]{Colussi_dynamical_2018}%
  \BibitemOpen
  \bibfield  {author} {\bibinfo {author} {\bibfnamefont {V.~E.}\ \bibnamefont
  {Colussi}}, \bibinfo {author} {\bibfnamefont {S.}~\bibnamefont {Musolino}}, \
  and\ \bibinfo {author} {\bibfnamefont {S.~J. J. M.~F.}\ \bibnamefont
  {Kokkelmans}},\ }\bibfield  {title} {\enquote {\bibinfo {title} {Dynamical
  formation of the unitary bose gas},}\ }\href@noop {} {\bibfield  {journal}
  {\bibinfo  {journal} {Phys. Rev. A}\ }\textbf {\bibinfo {volume} {98}},\
  \bibinfo {pages} {051601} (\bibinfo {year} {2018}{\natexlab{a}})}\BibitemShut
  {NoStop}%
\bibitem [{\citenamefont {Makotyn}\ \emph {et~al.}(2014)\citenamefont
  {Makotyn}, \citenamefont {Klauss}, \citenamefont {Goldberger}, \citenamefont
  {Cornell},\ and\ \citenamefont {Jin}}]{makotyn_universal_2014}%
  \BibitemOpen
  \bibfield  {author} {\bibinfo {author} {\bibfnamefont {P.}~\bibnamefont
  {Makotyn}}, \bibinfo {author} {\bibfnamefont {C.~E.}\ \bibnamefont {Klauss}},
  \bibinfo {author} {\bibfnamefont {D.~L.}\ \bibnamefont {Goldberger}},
  \bibinfo {author} {\bibfnamefont {E.~A.}\ \bibnamefont {Cornell}}, \ and\
  \bibinfo {author} {\bibfnamefont {D.~S.}\ \bibnamefont {Jin}},\ }\bibfield
  {title} {\enquote {\bibinfo {title} {Universal dynamics of a degenerate
  unitary bose gas},}\ }\href@noop {} {\bibfield  {journal} {\bibinfo
  {journal} {Nature Phys.}\ }\textbf {\bibinfo {volume} {10}},\ \bibinfo
  {pages} {116--119} (\bibinfo {year} {2014})}\BibitemShut {NoStop}%
\bibitem [{\citenamefont {Eigen}\ \emph {et~al.}(2018)\citenamefont {Eigen},
  \citenamefont {Glidden}, \citenamefont {Lopes}, \citenamefont {Cornell},
  \citenamefont {Smith},\ and\ \citenamefont
  {Hadzibabic}}]{eigen_universal_2018}%
  \BibitemOpen
  \bibfield  {author} {\bibinfo {author} {\bibfnamefont {C.}~\bibnamefont
  {Eigen}}, \bibinfo {author} {\bibfnamefont {J.~A.~P.}\ \bibnamefont
  {Glidden}}, \bibinfo {author} {\bibfnamefont {R.}~\bibnamefont {Lopes}},
  \bibinfo {author} {\bibfnamefont {E.~A.}\ \bibnamefont {Cornell}}, \bibinfo
  {author} {\bibfnamefont {R.~P.}\ \bibnamefont {Smith}}, \ and\ \bibinfo
  {author} {\bibfnamefont {Z.}~\bibnamefont {Hadzibabic}},\ }\bibfield  {title}
  {\enquote {\bibinfo {title} {Universal prethermal dynamics of bose gases
  quenched to unitarity},}\ }\href@noop {} {\bibfield  {journal} {\bibinfo
  {journal} {Nature}\ }\textbf {\bibinfo {volume} {563}},\ \bibinfo {pages}
  {221--224} (\bibinfo {year} {2018})}\BibitemShut {NoStop}%
\bibitem [{\citenamefont {Klauss}\ \emph {et~al.}(2017)\citenamefont {Klauss},
  \citenamefont {Xie}, \citenamefont {Lopez-Abadia}, \citenamefont {D’Incao},
  \citenamefont {Hadzibabic}, \citenamefont {Jin},\ and\ \citenamefont
  {Cornell}}]{klauss_observation_2017}%
  \BibitemOpen
  \bibfield  {author} {\bibinfo {author} {\bibfnamefont {C.~E.}\ \bibnamefont
  {Klauss}}, \bibinfo {author} {\bibfnamefont {X.}~\bibnamefont {Xie}},
  \bibinfo {author} {\bibfnamefont {C.}~\bibnamefont {Lopez-Abadia}}, \bibinfo
  {author} {\bibfnamefont {J.~P.}\ \bibnamefont {D’Incao}}, \bibinfo {author}
  {\bibfnamefont {Z.}~\bibnamefont {Hadzibabic}}, \bibinfo {author}
  {\bibfnamefont {D.~S.}\ \bibnamefont {Jin}}, \ and\ \bibinfo {author}
  {\bibfnamefont {E.~A.}\ \bibnamefont {Cornell}},\ }\bibfield  {title}
  {\enquote {\bibinfo {title} {Observation of efimov molecules created from a
  resonantly interacting bose gas},}\ }\href {\doibase
  10.1103/PhysRevLett.119.143401} {\bibfield  {journal} {\bibinfo  {journal}
  {Phys. Rev. Lett.}\ }\textbf {\bibinfo {volume} {119}},\ \bibinfo {pages}
  {143401} (\bibinfo {year} {2017})}\BibitemShut {NoStop}%
\bibitem [{\citenamefont {Fletcher}\ \emph {et~al.}(2017)\citenamefont
  {Fletcher}, \citenamefont {Lopes}, \citenamefont {Man}, \citenamefont
  {Navon}, \citenamefont {Smith}, \citenamefont {Zwierlein},\ and\
  \citenamefont {Hadzibabic}}]{fletcher_two-_2017}%
  \BibitemOpen
  \bibfield  {author} {\bibinfo {author} {\bibfnamefont {R.~J.}\ \bibnamefont
  {Fletcher}}, \bibinfo {author} {\bibfnamefont {R.}~\bibnamefont {Lopes}},
  \bibinfo {author} {\bibfnamefont {J.}~\bibnamefont {Man}}, \bibinfo {author}
  {\bibfnamefont {N.}~\bibnamefont {Navon}}, \bibinfo {author} {\bibfnamefont
  {R.~P.}\ \bibnamefont {Smith}}, \bibinfo {author} {\bibfnamefont {M.~W.}\
  \bibnamefont {Zwierlein}}, \ and\ \bibinfo {author} {\bibfnamefont
  {Z.}~\bibnamefont {Hadzibabic}},\ }\bibfield  {title} {\enquote {\bibinfo
  {title} {Two-and three-body contacts in the unitary bose gas},}\ }\href@noop
  {} {\bibfield  {journal} {\bibinfo  {journal} {Science}\ }\textbf {\bibinfo
  {volume} {355}},\ \bibinfo {pages} {377--380} (\bibinfo {year}
  {2017})}\BibitemShut {NoStop}%
\bibitem [{\citenamefont {Chen}\ \emph {et~al.}(2022)\citenamefont {Chen},
  \citenamefont {Duda}, \citenamefont {Schindewolf}, \citenamefont {Bause},
  \citenamefont {Bloch},\ and\ \citenamefont {Luo}}]{blochprl2022supress}%
  \BibitemOpen
  \bibfield  {author} {\bibinfo {author} {\bibfnamefont {X.-Y.}\ \bibnamefont
  {Chen}}, \bibinfo {author} {\bibfnamefont {M.}~\bibnamefont {Duda}}, \bibinfo
  {author} {\bibfnamefont {A.}~\bibnamefont {Schindewolf}}, \bibinfo {author}
  {\bibfnamefont {R.}~\bibnamefont {Bause}}, \bibinfo {author} {\bibfnamefont
  {I.}~\bibnamefont {Bloch}}, \ and\ \bibinfo {author} {\bibfnamefont {X.-Y.}\
  \bibnamefont {Luo}},\ }\bibfield  {title} {\enquote {\bibinfo {title}
  {Suppression of unitary three-body loss in a degenerate bose-fermi
  mixture},}\ }\href {\doibase 10.1103/PhysRevLett.128.153401} {\bibfield
  {journal} {\bibinfo  {journal} {Phys. Rev. Lett.}\ }\textbf {\bibinfo
  {volume} {128}},\ \bibinfo {pages} {153401} (\bibinfo {year}
  {2022})}\BibitemShut {NoStop}%
\bibitem [{\citenamefont {Donley}\ \emph {et~al.}(2002)\citenamefont {Donley},
  \citenamefont {Claussen}, \citenamefont {Thompson},\ and\ \citenamefont
  {Wieman}}]{donley_atommolecule_2002}%
  \BibitemOpen
  \bibfield  {author} {\bibinfo {author} {\bibfnamefont {E.~A.}\ \bibnamefont
  {Donley}}, \bibinfo {author} {\bibfnamefont {N.~R.}\ \bibnamefont
  {Claussen}}, \bibinfo {author} {\bibfnamefont {S.~T.}\ \bibnamefont
  {Thompson}}, \ and\ \bibinfo {author} {\bibfnamefont {C.~E.}\ \bibnamefont
  {Wieman}},\ }\bibfield  {title} {\enquote {\bibinfo {title} {Atom--molecule
  coherence in a bose--einstein condensate},}\ }\href {\doibase
  10.1038/417529a} {\bibfield  {journal} {\bibinfo  {journal} {Nature}\
  }\textbf {\bibinfo {volume} {417}},\ \bibinfo {pages} {529--533} (\bibinfo
  {year} {2002})}\BibitemShut {NoStop}%
\bibitem [{\citenamefont {Chin}\ \emph {et~al.}(2010)\citenamefont {Chin},
  \citenamefont {Grimm}, \citenamefont {Julienne},\ and\ \citenamefont
  {Tiesinga}}]{chin_feshbach_2010}%
  \BibitemOpen
  \bibfield  {author} {\bibinfo {author} {\bibfnamefont {C.}~\bibnamefont
  {Chin}}, \bibinfo {author} {\bibfnamefont {R.}~\bibnamefont {Grimm}},
  \bibinfo {author} {\bibfnamefont {P.}~\bibnamefont {Julienne}}, \ and\
  \bibinfo {author} {\bibfnamefont {E.}~\bibnamefont {Tiesinga}},\ }\bibfield
  {title} {\enquote {\bibinfo {title} {Feshbach resonances in ultracold
  gases},}\ }\href {\doibase 10.1103/RevModPhys.82.1225} {\bibfield  {journal}
  {\bibinfo  {journal} {Rev. Mod. Phys.}\ }\textbf {\bibinfo {volume} {82}},\
  \bibinfo {pages} {1225--1286} (\bibinfo {year} {2010})}\BibitemShut {NoStop}%
\bibitem [{\citenamefont {Yudkin}\ \emph {et~al.}(2019)\citenamefont {Yudkin},
  \citenamefont {Elbaz}, \citenamefont {Giannakeas}, \citenamefont {Greene},\
  and\ \citenamefont {Khaykovich}}]{yudkin_coherent_2019}%
  \BibitemOpen
  \bibfield  {author} {\bibinfo {author} {\bibfnamefont {Y.}~\bibnamefont
  {Yudkin}}, \bibinfo {author} {\bibfnamefont {R.}~\bibnamefont {Elbaz}},
  \bibinfo {author} {\bibfnamefont {P.}~\bibnamefont {Giannakeas}}, \bibinfo
  {author} {\bibfnamefont {C.~H.}\ \bibnamefont {Greene}}, \ and\ \bibinfo
  {author} {\bibfnamefont {L.}~\bibnamefont {Khaykovich}},\ }\bibfield  {title}
  {\enquote {\bibinfo {title} {Coherent superposition of feshbach dimers and
  efimov trimers},}\ }\href {\doibase 10.1103/PhysRevLett.122.200402}
  {\bibfield  {journal} {\bibinfo  {journal} {Phys. Rev. Lett.}\ }\textbf
  {\bibinfo {volume} {122}},\ \bibinfo {pages} {200402} (\bibinfo {year}
  {2019})}\BibitemShut {NoStop}%
\bibitem [{\citenamefont {Yudkin}\ \emph {et~al.}(2020)\citenamefont {Yudkin},
  \citenamefont {Elbaz},\ and\ \citenamefont
  {Khaykovich}}]{yudkin_efimov_2020}%
  \BibitemOpen
  \bibfield  {author} {\bibinfo {author} {\bibfnamefont {Y.}~\bibnamefont
  {Yudkin}}, \bibinfo {author} {\bibfnamefont {R.}~\bibnamefont {Elbaz}}, \
  and\ \bibinfo {author} {\bibfnamefont {L.}~\bibnamefont {Khaykovich}},\
  }\bibfield  {title} {\enquote {\bibinfo {title} {Efimov energy level
  rebounding off the atom-dimer continuum},}\ }\href
  {http://arxiv.org/abs/2004.02723} {\bibfield  {journal} {\bibinfo  {journal}
  {arXiv:2004.02723}\ } (\bibinfo {year} {2020})}\BibitemShut {NoStop}%
\bibitem [{\citenamefont {Góral}\ \emph {et~al.}(2004)\citenamefont {Góral},
  \citenamefont {Köhler}, \citenamefont {Gardiner}, \citenamefont {Tiesinga},\
  and\ \citenamefont {Julienne}}]{Goral_adiabatic_2004}%
  \BibitemOpen
  \bibfield  {author} {\bibinfo {author} {\bibfnamefont {K.}~\bibnamefont
  {Góral}}, \bibinfo {author} {\bibfnamefont {T.}~\bibnamefont {Köhler}},
  \bibinfo {author} {\bibfnamefont {S.~A.}\ \bibnamefont {Gardiner}}, \bibinfo
  {author} {\bibfnamefont {E.}~\bibnamefont {Tiesinga}}, \ and\ \bibinfo
  {author} {\bibfnamefont {P.~S.}\ \bibnamefont {Julienne}},\ }\bibfield
  {title} {\enquote {\bibinfo {title} {Adiabatic association of ultracold
  molecules via magnetic-field tunable interactions},}\ }\href {\doibase
  10.1088/0953-4075/37/17/006} {\bibfield  {journal} {\bibinfo  {journal} {J.
  Phys. B: At. Mol. Opt. Phys.}\ }\textbf {\bibinfo {volume} {37}},\ \bibinfo
  {pages} {3457} (\bibinfo {year} {2004})}\BibitemShut {NoStop}%
\bibitem [{\citenamefont {Sykes}\ \emph {et~al.}(2014)\citenamefont {Sykes},
  \citenamefont {Corson}, \citenamefont {D'Incao}, \citenamefont {Koller},
  \citenamefont {Greene}, \citenamefont {Rey}, \citenamefont {Hazzard},\ and\
  \citenamefont {Bohn}}]{sykes_quenching_2014}%
  \BibitemOpen
  \bibfield  {author} {\bibinfo {author} {\bibfnamefont {A.~G.}\ \bibnamefont
  {Sykes}}, \bibinfo {author} {\bibfnamefont {J.~P.}\ \bibnamefont {Corson}},
  \bibinfo {author} {\bibfnamefont {J.~P.}\ \bibnamefont {D'Incao}}, \bibinfo
  {author} {\bibfnamefont {A.~P.}\ \bibnamefont {Koller}}, \bibinfo {author}
  {\bibfnamefont {C.~H.}\ \bibnamefont {Greene}}, \bibinfo {author}
  {\bibfnamefont {A.~M.}\ \bibnamefont {Rey}}, \bibinfo {author} {\bibfnamefont
  {K.~R.~A.}\ \bibnamefont {Hazzard}}, \ and\ \bibinfo {author} {\bibfnamefont
  {J.~L.}\ \bibnamefont {Bohn}},\ }\bibfield  {title} {\enquote {\bibinfo
  {title} {Quenching to unitarity: Quantum dynamics in a three-dimensional bose
  gas},}\ }\href {\doibase 10.1103/PhysRevA.89.021601} {\bibfield  {journal}
  {\bibinfo  {journal} {Phys. Rev. A}\ }\textbf {\bibinfo {volume} {89}},\
  \bibinfo {pages} {021601} (\bibinfo {year} {2014})}\BibitemShut {NoStop}%
\bibitem [{\citenamefont {Corson}\ and\ \citenamefont
  {Bohn}(2015)}]{corson_bound-state_2015}%
  \BibitemOpen
  \bibfield  {author} {\bibinfo {author} {\bibfnamefont {J.~P.}\ \bibnamefont
  {Corson}}\ and\ \bibinfo {author} {\bibfnamefont {J.~L.}\ \bibnamefont
  {Bohn}},\ }\bibfield  {title} {\enquote {\bibinfo {title} {Bound-state
  signatures in quenched bose-einstein condensates},}\ }\href {\doibase
  10.1103/PhysRevA.91.013616} {\bibfield  {journal} {\bibinfo  {journal} {Phys.
  Rev. A}\ }\textbf {\bibinfo {volume} {91}},\ \bibinfo {pages} {013616}
  (\bibinfo {year} {2015})}\BibitemShut {NoStop}%
\bibitem [{\citenamefont {Borca}\ \emph {et~al.}(2003)\citenamefont {Borca},
  \citenamefont {Blume},\ and\ \citenamefont {Greene}}]{borca_two-atom_2003}%
  \BibitemOpen
  \bibfield  {author} {\bibinfo {author} {\bibfnamefont {B.}~\bibnamefont
  {Borca}}, \bibinfo {author} {\bibfnamefont {D.}~\bibnamefont {Blume}}, \ and\
  \bibinfo {author} {\bibfnamefont {C.~H.}\ \bibnamefont {Greene}},\ }\bibfield
   {title} {\enquote {\bibinfo {title} {A two-atom picture of coherent
  atom--molecule quantum beats},}\ }\href {\doibase 10.1088/1367-2630/5/1/111}
  {\bibfield  {journal} {\bibinfo  {journal} {New J. Phys.}\ }\textbf {\bibinfo
  {volume} {5}},\ \bibinfo {pages} {111} (\bibinfo {year} {2003})}\BibitemShut
  {NoStop}%
\bibitem [{\citenamefont {D’Incao}\ \emph {et~al.}(2018)\citenamefont
  {D’Incao}, \citenamefont {Wang},\ and\ \citenamefont
  {Colussi}}]{dincao_efimov_2018}%
  \BibitemOpen
  \bibfield  {author} {\bibinfo {author} {\bibfnamefont {J.~P.}\ \bibnamefont
  {D’Incao}}, \bibinfo {author} {\bibfnamefont {J.}~\bibnamefont {Wang}}, \
  and\ \bibinfo {author} {\bibfnamefont {V.~E.}\ \bibnamefont {Colussi}},\
  }\bibfield  {title} {\enquote {\bibinfo {title} {Efimov physics in quenched
  unitary bose gases},}\ }\href {\doibase 10.1103/PhysRevLett.121.023401}
  {\bibfield  {journal} {\bibinfo  {journal} {Phys. Rev. Lett.}\ }\textbf
  {\bibinfo {volume} {121}},\ \bibinfo {pages} {023401} (\bibinfo {year}
  {2018})}\BibitemShut {NoStop}%
\bibitem [{\citenamefont {von Stecher}\ and\ \citenamefont
  {Greene}(2007)}]{von_stecher_spectrum_2007}%
  \BibitemOpen
  \bibfield  {author} {\bibinfo {author} {\bibfnamefont {J.}~\bibnamefont {von
  Stecher}}\ and\ \bibinfo {author} {\bibfnamefont {C.~H.}\ \bibnamefont
  {Greene}},\ }\bibfield  {title} {\enquote {\bibinfo {title} {Spectrum and
  dynamics of the bcs-bec crossover from a few-body perspective},}\ }\href
  {\doibase 10.1103/PhysRevLett.99.090402} {\bibfield  {journal} {\bibinfo
  {journal} {Phys. Rev. Lett.}\ }\textbf {\bibinfo {volume} {99}},\ \bibinfo
  {pages} {090402} (\bibinfo {year} {2007})}\BibitemShut {NoStop}%
\bibitem [{\citenamefont {Rittenhouse}\ \emph {et~al.}(2010)\citenamefont
  {Rittenhouse}, \citenamefont {Mehta},\ and\ \citenamefont
  {Greene}}]{rittenhouse_greens_2010}%
  \BibitemOpen
  \bibfield  {author} {\bibinfo {author} {\bibfnamefont {S.~T.}\ \bibnamefont
  {Rittenhouse}}, \bibinfo {author} {\bibfnamefont {N.~P.}\ \bibnamefont
  {Mehta}}, \ and\ \bibinfo {author} {\bibfnamefont {C.~H.}\ \bibnamefont
  {Greene}},\ }\bibfield  {title} {\enquote {\bibinfo {title} {Green’s
  functions and the adiabatic hyperspherical method},}\ }\href {\doibase
  10.1103/PhysRevA.82.022706} {\bibfield  {journal} {\bibinfo  {journal} {Phys.
  Rev. A}\ }\textbf {\bibinfo {volume} {82}},\ \bibinfo {pages} {022706}
  (\bibinfo {year} {2010})}\BibitemShut {NoStop}%
\bibitem [{\citenamefont {Burstein}\ and\ \citenamefont
  {Mirin}(1970)}]{Burstein_third_1970}%
  \BibitemOpen
  \bibfield  {author} {\bibinfo {author} {\bibfnamefont {S.~Z.}\ \bibnamefont
  {Burstein}}\ and\ \bibinfo {author} {\bibfnamefont {A.~A.}\ \bibnamefont
  {Mirin}},\ }\bibfield  {title} {\enquote {\bibinfo {title} {Third order
  difference methods for hyperbolic equations},}\ }\href {\doibase
  https://doi.org/10.1016/0021-9991(70)90080-X} {\bibfield  {journal} {\bibinfo
   {journal} {J. Comput. Phys.}\ }\textbf {\bibinfo {volume} {5}},\ \bibinfo
  {pages} {547--571} (\bibinfo {year} {1970})}\BibitemShut {NoStop}%
\bibitem [{\citenamefont {Tarana}\ and\ \citenamefont
  {Greene}(2012)}]{Tarana_femtosecond_2012}%
  \BibitemOpen
  \bibfield  {author} {\bibinfo {author} {\bibfnamefont {M.}~\bibnamefont
  {Tarana}}\ and\ \bibinfo {author} {\bibfnamefont {C.~H.}\ \bibnamefont
  {Greene}},\ }\bibfield  {title} {\enquote {\bibinfo {title} {Femtosecond
  transparency in the extreme-ultraviolet region},}\ }\href@noop {} {\bibfield
  {journal} {\bibinfo  {journal} {Phys. Rev. A}\ }\textbf {\bibinfo {volume}
  {85}},\ \bibinfo {pages} {013411} (\bibinfo {year} {2012})}\BibitemShut
  {NoStop}%
\bibitem [{\citenamefont {Giannakeas}\ \emph {et~al.}(2019)\citenamefont
  {Giannakeas}, \citenamefont {Khaykovich}, \citenamefont {Rost},\ and\
  \citenamefont {Greene}}]{giannakeas_nonadiabatic_2019}%
  \BibitemOpen
  \bibfield  {author} {\bibinfo {author} {\bibfnamefont {P.}~\bibnamefont
  {Giannakeas}}, \bibinfo {author} {\bibfnamefont {L.}~\bibnamefont
  {Khaykovich}}, \bibinfo {author} {\bibfnamefont {J.-M.}\ \bibnamefont
  {Rost}}, \ and\ \bibinfo {author} {\bibfnamefont {C.~H.}\ \bibnamefont
  {Greene}},\ }\bibfield  {title} {\enquote {\bibinfo {title} {Nonadiabatic
  molecular association in thermal gases driven by radio-frequency pulses},}\
  }\href@noop {} {\bibfield  {journal} {\bibinfo  {journal} {Phys. Rev. Lett.}\
  }\textbf {\bibinfo {volume} {123}},\ \bibinfo {pages} {043204} (\bibinfo
  {year} {2019})}\BibitemShut {NoStop}%
\bibitem [{\citenamefont {Lambropoulos}\ and\ \citenamefont
  {Petrosyan}(2006)}]{lambropoulos2007fundamentals}%
  \BibitemOpen
  \bibfield  {author} {\bibinfo {author} {\bibfnamefont {P.}~\bibnamefont
  {Lambropoulos}}\ and\ \bibinfo {author} {\bibfnamefont {D.}~\bibnamefont
  {Petrosyan}},\ }\href@noop {} {\emph {\bibinfo {title} {Fundamentals of
  Quantum Optics and Quantum Information}}}\ (\bibinfo  {publisher}
  {Springer-Verlag},\ \bibinfo {address} {Berlin, Heidelberg},\ \bibinfo {year}
  {2006})\BibitemShut {NoStop}%
\bibitem [{\citenamefont {Werner}\ and\ \citenamefont
  {Castin}(2006)}]{werner_unitary_2006}%
  \BibitemOpen
  \bibfield  {author} {\bibinfo {author} {\bibfnamefont {F.}~\bibnamefont
  {Werner}}\ and\ \bibinfo {author} {\bibfnamefont {Y.}~\bibnamefont
  {Castin}},\ }\bibfield  {title} {\enquote {\bibinfo {title} {Unitary quantum
  three-body problem in a harmonic trap},}\ }\href {\doibase
  10.1103/PhysRevLett.97.150401} {\bibfield  {journal} {\bibinfo  {journal}
  {Phys. Rev. Lett.}\ }\textbf {\bibinfo {volume} {97}},\ \bibinfo {pages}
  {150401} (\bibinfo {year} {2006})}\BibitemShut {NoStop}%
\bibitem [{\citenamefont {Petrov}\ \emph {et~al.}(2004)\citenamefont {Petrov},
  \citenamefont {Salomon},\ and\ \citenamefont
  {Shlyapnikov}}]{petrov_weakly_2004}%
  \BibitemOpen
  \bibfield  {author} {\bibinfo {author} {\bibfnamefont {D.~S.}\ \bibnamefont
  {Petrov}}, \bibinfo {author} {\bibfnamefont {C.}~\bibnamefont {Salomon}}, \
  and\ \bibinfo {author} {\bibfnamefont {G.~V.}\ \bibnamefont {Shlyapnikov}},\
  }\bibfield  {title} {\enquote {\bibinfo {title} {Weakly bound dimers of
  fermionic atoms},}\ }\href {\doibase 10.1103/PhysRevLett.93.090404}
  {\bibfield  {journal} {\bibinfo  {journal} {Phys. Rev. Lett.}\ }\textbf
  {\bibinfo {volume} {93}},\ \bibinfo {pages} {090404} (\bibinfo {year}
  {2004})}\BibitemShut {NoStop}%
\bibitem [{\citenamefont {Nielsen}\ \emph {et~al.}(2002)\citenamefont
  {Nielsen}, \citenamefont {Suno},\ and\ \citenamefont
  {Esry}}]{nielsen_efimov_2002}%
  \BibitemOpen
  \bibfield  {author} {\bibinfo {author} {\bibfnamefont {E.}~\bibnamefont
  {Nielsen}}, \bibinfo {author} {\bibfnamefont {H.}~\bibnamefont {Suno}}, \
  and\ \bibinfo {author} {\bibfnamefont {B.~D.}\ \bibnamefont {Esry}},\
  }\bibfield  {title} {\enquote {\bibinfo {title} {Efimov resonances in
  atom-diatom scattering},}\ }\href {\doibase 10.1103/PhysRevA.66.012705}
  {\bibfield  {journal} {\bibinfo  {journal} {Phys. Rev. A}\ }\textbf {\bibinfo
  {volume} {66}},\ \bibinfo {pages} {012705} (\bibinfo {year}
  {2002})}\BibitemShut {NoStop}%
\bibitem [{\citenamefont {Braaten}\ and\ \citenamefont
  {Hammer}(2004)}]{Braaten_enhanced_2004}%
  \BibitemOpen
  \bibfield  {author} {\bibinfo {author} {\bibfnamefont {E.}~\bibnamefont
  {Braaten}}\ and\ \bibinfo {author} {\bibfnamefont {H.-W.}\ \bibnamefont
  {Hammer}},\ }\bibfield  {title} {\enquote {\bibinfo {title} {Enhanced dimer
  relaxation in an atomic and molecular bose-einstein condensate},}\ }\href
  {\doibase 10.1103/PhysRevA.70.042706} {\bibfield  {journal} {\bibinfo
  {journal} {Phys. Rev. A}\ }\textbf {\bibinfo {volume} {70}},\ \bibinfo
  {pages} {042706} (\bibinfo {year} {2004})}\BibitemShut {NoStop}%
\bibitem [{\citenamefont {Claussen}\ \emph {et~al.}(2003)\citenamefont
  {Claussen}, \citenamefont {Kokkelmans}, \citenamefont {Thompson},
  \citenamefont {Donley}, \citenamefont {Hodby},\ and\ \citenamefont
  {Wieman}}]{Claussen_precision_2003}%
  \BibitemOpen
  \bibfield  {author} {\bibinfo {author} {\bibfnamefont {N.~R.}\ \bibnamefont
  {Claussen}}, \bibinfo {author} {\bibfnamefont {S.~J. J. M.~F.}\ \bibnamefont
  {Kokkelmans}}, \bibinfo {author} {\bibfnamefont {S.~T.}\ \bibnamefont
  {Thompson}}, \bibinfo {author} {\bibfnamefont {E.~A.}\ \bibnamefont
  {Donley}}, \bibinfo {author} {\bibfnamefont {E.}~\bibnamefont {Hodby}}, \
  and\ \bibinfo {author} {\bibfnamefont {C.~E.}\ \bibnamefont {Wieman}},\
  }\bibfield  {title} {\enquote {\bibinfo {title} {Very-high-precision
  bound-state spectroscopy near a 85 rb feshbach resonance},}\ }\href {\doibase
  10.1103/PhysRevA.67.060701} {\bibfield  {journal} {\bibinfo  {journal} {Phys.
  Rev. A}\ }\textbf {\bibinfo {volume} {67}},\ \bibinfo {pages} {060701}
  (\bibinfo {year} {2003})}\BibitemShut {NoStop}%
\bibitem [{\citenamefont {K\"ohler}\ \emph {et~al.}(2005)\citenamefont
  {K\"ohler}, \citenamefont {Tiesinga},\ and\ \citenamefont
  {Julienne}}]{Kohler_spontaneous_2005}%
  \BibitemOpen
  \bibfield  {author} {\bibinfo {author} {\bibfnamefont {Th.}\ \bibnamefont
  {K\"ohler}}, \bibinfo {author} {\bibfnamefont {E.}~\bibnamefont {Tiesinga}},
  \ and\ \bibinfo {author} {\bibfnamefont {P.~S.}\ \bibnamefont {Julienne}},\
  }\bibfield  {title} {\enquote {\bibinfo {title} {Spontaneous dissociation of
  long-range feshbach molecules},}\ }\href {\doibase
  10.1103/PhysRevLett.94.020402} {\bibfield  {journal} {\bibinfo  {journal}
  {Phys. Rev. Lett.}\ }\textbf {\bibinfo {volume} {94}},\ \bibinfo {pages}
  {020402} (\bibinfo {year} {2005})}\BibitemShut {NoStop}%
\bibitem [{\citenamefont {Colussi}\ \emph
  {et~al.}(2018{\natexlab{b}})\citenamefont {Colussi}, \citenamefont {Corson},\
  and\ \citenamefont {D’Incao}}]{colussi_dynamics_2018}%
  \BibitemOpen
  \bibfield  {author} {\bibinfo {author} {\bibfnamefont {V.~E.}\ \bibnamefont
  {Colussi}}, \bibinfo {author} {\bibfnamefont {J.~P.}\ \bibnamefont {Corson}},
  \ and\ \bibinfo {author} {\bibfnamefont {J.~P.}\ \bibnamefont {D’Incao}},\
  }\bibfield  {title} {\enquote {\bibinfo {title} {Dynamics of three-body
  correlations in quenched unitary bose gases},}\ }\href {\doibase
  10.1103/PhysRevLett.120.100401} {\bibfield  {journal} {\bibinfo  {journal}
  {Phys. Rev. Lett.}\ }\textbf {\bibinfo {volume} {120}},\ \bibinfo {pages}
  {100401} (\bibinfo {year} {2018}{\natexlab{b}})}\BibitemShut {NoStop}%
\bibitem [{\citenamefont {Colussi}\ \emph {et~al.}(2019)\citenamefont
  {Colussi}, \citenamefont {van Zwol}, \citenamefont {D'Incao},\ and\
  \citenamefont {Kokkelmans}}]{colussi_bunching_2019}%
  \BibitemOpen
  \bibfield  {author} {\bibinfo {author} {\bibfnamefont {V.~E.}\ \bibnamefont
  {Colussi}}, \bibinfo {author} {\bibfnamefont {B.~E.}\ \bibnamefont {van
  Zwol}}, \bibinfo {author} {\bibfnamefont {J.~P.}\ \bibnamefont {D'Incao}}, \
  and\ \bibinfo {author} {\bibfnamefont {S.~J. J. M.~F.}\ \bibnamefont
  {Kokkelmans}},\ }\bibfield  {title} {\enquote {\bibinfo {title} {Bunching,
  clustering, and the buildup of few-body correlations in a quenched unitary
  bose gas},}\ }\href@noop {} {\bibfield  {journal} {\bibinfo  {journal} {Phys.
  Rev. A}\ }\textbf {\bibinfo {volume} {99}},\ \bibinfo {pages} {043604}
  (\bibinfo {year} {2019})}\BibitemShut {NoStop}%
\bibitem [{\citenamefont {Braaten}\ and\ \citenamefont
  {Hammer}(2006)}]{braaten_universality_2006}%
  \BibitemOpen
  \bibfield  {author} {\bibinfo {author} {\bibfnamefont {E.}~\bibnamefont
  {Braaten}}\ and\ \bibinfo {author} {\bibfnamefont {H.~W.}\ \bibnamefont
  {Hammer}},\ }\bibfield  {title} {\enquote {\bibinfo {title} {Universality in
  few-body systems with large scattering length},}\ }\href {\doibase
  10.1016/j.physrep.2006.03.001} {\bibfield  {journal} {\bibinfo  {journal}
  {Phys. Rep.}\ }\textbf {\bibinfo {volume} {428}},\ \bibinfo {pages}
  {259--390} (\bibinfo {year} {2006})}\BibitemShut {NoStop}%
\bibitem [{\citenamefont {Yudkin}\ \emph {et~al.}(2023)\citenamefont {Yudkin},
  \citenamefont {Elbaz}, \citenamefont {D'Incao}, \citenamefont {Julienne},\
  and\ \citenamefont {Khaykovich}}]{yudkin_reshape_2023}%
  \BibitemOpen
  \bibfield  {author} {\bibinfo {author} {\bibfnamefont {Y.}~\bibnamefont
  {Yudkin}}, \bibinfo {author} {\bibfnamefont {R.}~\bibnamefont {Elbaz}},
  \bibinfo {author} {\bibfnamefont {J.~P.}\ \bibnamefont {D'Incao}}, \bibinfo
  {author} {\bibfnamefont {P.~S.}\ \bibnamefont {Julienne}}, \ and\ \bibinfo
  {author} {\bibfnamefont {L.}~\bibnamefont {Khaykovich}},\ }\href@noop {}
  {\enquote {\bibinfo {title} {The reshape of three-body interactions:
  Observation of the survival of an efimov state in the atom-dimer
  continuum},}\ } (\bibinfo {year} {2023}),\ \Eprint
  {http://arxiv.org/abs/2308.06237} {arXiv:2308.06237 [cond-mat.quant-gas]}
  \BibitemShut {NoStop}%
\bibitem [{\citenamefont {Etrych}\ \emph {et~al.}(2023)\citenamefont {Etrych},
  \citenamefont {Martirosyan}, \citenamefont {Cao}, \citenamefont {Glidden},
  \citenamefont {Dogra}, \citenamefont {Hutson}, \citenamefont {Hadzibabic},\
  and\ \citenamefont {Eigen}}]{Etrych_pinpointing_2023}%
  \BibitemOpen
  \bibfield  {author} {\bibinfo {author} {\bibfnamefont {J.}~\bibnamefont
  {Etrych}}, \bibinfo {author} {\bibfnamefont {G.}~\bibnamefont {Martirosyan}},
  \bibinfo {author} {\bibfnamefont {A.}~\bibnamefont {Cao}}, \bibinfo {author}
  {\bibfnamefont {J.~A.~P.}\ \bibnamefont {Glidden}}, \bibinfo {author}
  {\bibfnamefont {L.~H.}\ \bibnamefont {Dogra}}, \bibinfo {author}
  {\bibfnamefont {J.~M.}\ \bibnamefont {Hutson}}, \bibinfo {author}
  {\bibfnamefont {Z.}~\bibnamefont {Hadzibabic}}, \ and\ \bibinfo {author}
  {\bibfnamefont {C.}~\bibnamefont {Eigen}},\ }\bibfield  {title} {\enquote
  {\bibinfo {title} {Pinpointing feshbach resonances and testing efimov
  universalities in $^{39}\mathrm{K}$},}\ }\href@noop {} {\bibfield  {journal}
  {\bibinfo  {journal} {Phys. Rev. Res.}\ }\textbf {\bibinfo {volume} {5}},\
  \bibinfo {pages} {013174} (\bibinfo {year} {2023})}\BibitemShut {NoStop}%
\bibitem [{\citenamefont {Xie}\ \emph {et~al.}(2020)\citenamefont {Xie},
  \citenamefont {Van~de Graaff}, \citenamefont {Chapurin}, \citenamefont
  {Frye}, \citenamefont {Hutson}, \citenamefont {D'Incao}, \citenamefont
  {Julienne}, \citenamefont {Ye},\ and\ \citenamefont
  {Cornell}}]{Xie_observation_2020}%
  \BibitemOpen
  \bibfield  {author} {\bibinfo {author} {\bibfnamefont {X.}~\bibnamefont
  {Xie}}, \bibinfo {author} {\bibfnamefont {M.~J.}\ \bibnamefont {Van~de
  Graaff}}, \bibinfo {author} {\bibfnamefont {R.}~\bibnamefont {Chapurin}},
  \bibinfo {author} {\bibfnamefont {M.~D.}\ \bibnamefont {Frye}}, \bibinfo
  {author} {\bibfnamefont {J.~M.}\ \bibnamefont {Hutson}}, \bibinfo {author}
  {\bibfnamefont {J.~P.}\ \bibnamefont {D'Incao}}, \bibinfo {author}
  {\bibfnamefont {P.~S.}\ \bibnamefont {Julienne}}, \bibinfo {author}
  {\bibfnamefont {J.}~\bibnamefont {Ye}}, \ and\ \bibinfo {author}
  {\bibfnamefont {E.~A.}\ \bibnamefont {Cornell}},\ }\bibfield  {title}
  {\enquote {\bibinfo {title} {Observation of efimov universality across a
  nonuniversal feshbach resonance in $^{39}\mathrm{K}$},}\ }\href@noop {}
  {\bibfield  {journal} {\bibinfo  {journal} {Phys. Rev. Lett.}\ }\textbf
  {\bibinfo {volume} {125}},\ \bibinfo {pages} {243401} (\bibinfo {year}
  {2020})}\BibitemShut {NoStop}%
\bibitem [{\citenamefont {Berninger}\ \emph {et~al.}(2011)\citenamefont
  {Berninger}, \citenamefont {Zenesini}, \citenamefont {Huang}, \citenamefont
  {Harm}, \citenamefont {N\"agerl}, \citenamefont {Ferlaino}, \citenamefont
  {Grimm}, \citenamefont {Julienne},\ and\ \citenamefont
  {Hutson}}]{Berninger_universality_2011}%
  \BibitemOpen
  \bibfield  {author} {\bibinfo {author} {\bibfnamefont {M.}~\bibnamefont
  {Berninger}}, \bibinfo {author} {\bibfnamefont {A.}~\bibnamefont {Zenesini}},
  \bibinfo {author} {\bibfnamefont {B.}~\bibnamefont {Huang}}, \bibinfo
  {author} {\bibfnamefont {W.}~\bibnamefont {Harm}}, \bibinfo {author}
  {\bibfnamefont {H.-C.}\ \bibnamefont {N\"agerl}}, \bibinfo {author}
  {\bibfnamefont {F.}~\bibnamefont {Ferlaino}}, \bibinfo {author}
  {\bibfnamefont {R.}~\bibnamefont {Grimm}}, \bibinfo {author} {\bibfnamefont
  {P.~S.}\ \bibnamefont {Julienne}}, \ and\ \bibinfo {author} {\bibfnamefont
  {J.~M.}\ \bibnamefont {Hutson}},\ }\bibfield  {title} {\enquote {\bibinfo
  {title} {Universality of the three-body parameter for efimov states in
  ultracold cesium},}\ }\href@noop {} {\bibfield  {journal} {\bibinfo
  {journal} {Phys. Rev. Lett.}\ }\textbf {\bibinfo {volume} {107}},\ \bibinfo
  {pages} {120401} (\bibinfo {year} {2011})}\BibitemShut {NoStop}%
\bibitem [{\citenamefont {Johansen}\ \emph {et~al.}(2017)\citenamefont
  {Johansen}, \citenamefont {DeSalvo}, \citenamefont {Patel},\ and\
  \citenamefont {Chin}}]{Johansen_testing_2017}%
  \BibitemOpen
  \bibfield  {author} {\bibinfo {author} {\bibfnamefont {J.}~\bibnamefont
  {Johansen}}, \bibinfo {author} {\bibfnamefont {B.~J.}\ \bibnamefont
  {DeSalvo}}, \bibinfo {author} {\bibfnamefont {K.}~\bibnamefont {Patel}}, \
  and\ \bibinfo {author} {\bibfnamefont {C.}~\bibnamefont {Chin}},\ }\bibfield
  {title} {\enquote {\bibinfo {title} {Testing universality of efimov physics
  across broad and narrow feshbach resonances},}\ }\href@noop {} {\bibfield
  {journal} {\bibinfo  {journal} {Nature Phys.}\ }\textbf {\bibinfo {volume}
  {13}},\ \bibinfo {pages} {731--735} (\bibinfo {year} {2017})}\BibitemShut
  {NoStop}%
\bibitem [{\citenamefont {Naidon}\ \emph {et~al.}(2014)\citenamefont {Naidon},
  \citenamefont {Endo},\ and\ \citenamefont {Ueda}}]{Naidon_microscopic_2014}%
  \BibitemOpen
  \bibfield  {author} {\bibinfo {author} {\bibfnamefont {P.}~\bibnamefont
  {Naidon}}, \bibinfo {author} {\bibfnamefont {S.}~\bibnamefont {Endo}}, \ and\
  \bibinfo {author} {\bibfnamefont {M.}~\bibnamefont {Ueda}},\ }\bibfield
  {title} {\enquote {\bibinfo {title} {Microscopic origin and universality
  classes of the efimov three-body parameter},}\ }\href@noop {} {\bibfield
  {journal} {\bibinfo  {journal} {Phys. Rev. Lett.}\ }\textbf {\bibinfo
  {volume} {112}},\ \bibinfo {pages} {105301} (\bibinfo {year}
  {2014})}\BibitemShut {NoStop}%
\bibitem [{\citenamefont {Wang}\ \emph {et~al.}(2012)\citenamefont {Wang},
  \citenamefont {D'Incao}, \citenamefont {Esry},\ and\ \citenamefont
  {Greene}}]{Wang_origin_2012}%
  \BibitemOpen
  \bibfield  {author} {\bibinfo {author} {\bibfnamefont {J.}~\bibnamefont
  {Wang}}, \bibinfo {author} {\bibfnamefont {J.~P.}\ \bibnamefont {D'Incao}},
  \bibinfo {author} {\bibfnamefont {B.~D.}\ \bibnamefont {Esry}}, \ and\
  \bibinfo {author} {\bibfnamefont {C.~H.}\ \bibnamefont {Greene}},\ }\bibfield
   {title} {\enquote {\bibinfo {title} {Origin of the three-body parameter
  universality in efimov physics},}\ }\href {\doibase
  10.1103/PhysRevLett.108.263001} {\bibfield  {journal} {\bibinfo  {journal}
  {Phys. Rev. Lett.}\ }\textbf {\bibinfo {volume} {108}},\ \bibinfo {pages}
  {263001} (\bibinfo {year} {2012})}\BibitemShut {NoStop}%
\bibitem [{\citenamefont {Bougas}\ \emph {et~al.}(2021)\citenamefont {Bougas},
  \citenamefont {Mistakidis}, \citenamefont {Giannakeas},\ and\ \citenamefont
  {Schmelcher}}]{bougas2021few}%
  \BibitemOpen
  \bibfield  {author} {\bibinfo {author} {\bibfnamefont {G.}~\bibnamefont
  {Bougas}}, \bibinfo {author} {\bibfnamefont {S.~I.}\ \bibnamefont
  {Mistakidis}}, \bibinfo {author} {\bibfnamefont {P.}~\bibnamefont
  {Giannakeas}}, \ and\ \bibinfo {author} {\bibfnamefont {P.}~\bibnamefont
  {Schmelcher}},\ }\bibfield  {title} {\enquote {\bibinfo {title} {Few-body
  correlations in two-dimensional bose and fermi ultracold mixtures},}\
  }\href@noop {} {\bibfield  {journal} {\bibinfo  {journal} {New J. Phys.}\
  }\textbf {\bibinfo {volume} {23}},\ \bibinfo {pages} {093022} (\bibinfo
  {year} {2021})}\BibitemShut {NoStop}%
\bibitem [{\citenamefont {Avery}(1989)}]{avery_hyperspherical_1989}%
  \BibitemOpen
  \bibfield  {author} {\bibinfo {author} {\bibfnamefont {J.}~\bibnamefont
  {Avery}},\ }\href@noop {} {\emph {\bibinfo {title} {Hyperspherical
  {Harmonics}: {Applications} in {Quantum} {Theory}}}}\ (\bibinfo  {publisher}
  {Kluwer Academic Publishers},\ \bibinfo {address} {Norwell, MA},\ \bibinfo
  {year} {1989})\BibitemShut {NoStop}%
\bibitem [{\citenamefont {Feynman}(1939)}]{Feynman_forces_1939}%
  \BibitemOpen
  \bibfield  {author} {\bibinfo {author} {\bibfnamefont {R.~P.}\ \bibnamefont
  {Feynman}},\ }\bibfield  {title} {\enquote {\bibinfo {title} {Forces in
  molecules},}\ }\href@noop {} {\bibfield  {journal} {\bibinfo  {journal}
  {Phys. Rev.}\ }\textbf {\bibinfo {volume} {56}},\ \bibinfo {pages} {340--343}
  (\bibinfo {year} {1939})}\BibitemShut {NoStop}%
\bibitem [{\citenamefont {Sakurai}(1967)}]{sakurai1967advanced}%
  \BibitemOpen
  \bibfield  {author} {\bibinfo {author} {\bibfnamefont {J.~J.}\ \bibnamefont
  {Sakurai}},\ }\href@noop {} {\emph {\bibinfo {title} {Advanced quantum
  mechanics}}}\ (\bibinfo  {publisher} {Pearson Education India},\ \bibinfo
  {year} {1967})\BibitemShut {NoStop}%
\bibitem [{\citenamefont {Gradshteyn}\ \emph {et~al.}(2015)\citenamefont
  {Gradshteyn}, \citenamefont {Ryzhik}, \citenamefont {Zwillinger},\ and\
  \citenamefont {Moll}}]{Gradshteyn_table_2015}%
  \BibitemOpen
  \bibfield  {author} {\bibinfo {author} {\bibfnamefont {I.~S.}\ \bibnamefont
  {Gradshteyn}}, \bibinfo {author} {\bibfnamefont {I.~M.}\ \bibnamefont
  {Ryzhik}}, \bibinfo {author} {\bibfnamefont {D.}~\bibnamefont {Zwillinger}},
  \ and\ \bibinfo {author} {\bibfnamefont {V.}~\bibnamefont {Moll}},\ }\href
  {\doibase 0123849330} {\emph {\bibinfo {title} {{Table of integrals, series,
  and products; 8th ed.}}}}\ (\bibinfo  {publisher} {Academic Press},\ \bibinfo
  {address} {Amsterdam},\ \bibinfo {year} {2015})\BibitemShut {NoStop}%
\end{thebibliography}%

\end{document}